\pdfoutput=1
\documentclass[preprint]{elsarticle}
\usepackage{amsmath,amssymb,latexsym}
\usepackage{graphicx}
\usepackage{bm}

\newcommand{\field}[1]{\mathbb{#1}}
\newcommand{\R}{\field{R}}

\newcommand{\PP}{{\boldsymbol P}}
\newcommand{\QQ}{{\boldsymbol Q}}

\begin{document} 

\begin{frontmatter}

\title{An adaptive algorithm for the cornea modeling from keratometric data}

\author{Andrei Mart\'{\i}nez-Finkelshtein\corref{cor1}}
\ead{andrei@ual.es}
\address{Department of Statistics and Applied Mathematics,
University of Almer\'{\i}a, Spain, and\\
Instituto Carlos I de F\'{\i}sica Te\'{o}rica y Computacional,
Granada University, Spain}

\author{Dar\'{\i}o Ramos-L\'opez}%
\address{%
 Department of Statistics and Applied Mathematics, University of Almer\'{\i}a, Spain}%

\author{Gracia M.~Castro-Luna} 
\address{
 Corporaci\'on VISSUM Almer\'{\i}a, Spain}%
 
\author{Jorge L.~Ali\'o}
\address{%
Corporaci\'on VISSUM Alicante, Spain
}%

\cortext[cor1]{Corresponding author. Send correspondence to Department of Statistics and Applied Mathematics,
University of Almer\'{\i}a, 04120 Almer\'{\i}a, Spain. Phone +34-950015217, fax +34-950015167.}

\date{\today}

\begin{abstract}
In this paper we describe an adaptive and multi-scale algorithm for the parsimonious fit of the corneal surface data that allows to adapt the number of functions used in the reconstruction to the conditions of each cornea. The method implements also a dynamical selection of the parameters and the management of noise. It can be used for the real-time reconstruction of both altimetric data and corneal power maps from the data collected by keratoscopes, such as the Placido rings based topographers, decisive for an early detection of corneal diseases such as keratoconus.

Numerical experiments show that the algorithm exhibits a steady exponential error decay, independently of the level of aberration of the cornea. The complexity of each anisotropic gaussian basis functions in the functional representation is the same, but their parameters vary to fit the current scale. This scale is determined only by the residual errors and not by the number of the iteration. Finally, the position and clustering of their centers, as well as the size of the shape parameters, provides an additional spatial information about the regions of higher irregularity. These results are compared with the standard approximation procedures based on the Zernike polynomials expansions.

\end{abstract} 


\end{frontmatter}



\section{\label{sec:level1}Introduction}

There is an increasing need of a reliable and precise modeling of corneal surfaces, motivated both by technological and clinical applications. 

Given the significance of the shape of the front surface of the cornea to the refraction of the eye \cite{Schwiegerling1989} and the ability to correct refractive errors by laser ablation of the front surface of the cornea, a detailed wavefront error analysis of individual corneal topography data is crucial for the clinicians as a basis for a customized treatment. It has been recognized that the corneal front surface generally provides the bulk of the ocular aberrations in the postsurgical or pathologic eye \cite{Artal2002}.

Corneal modeling can be used also as a tool for screening corneal diseases. Keratoconus, for example, distorts the corneal shape and results in a significant vision loss. Keratoconic patients should be screened for refractive surgeries because such treatments may worsen the corneal shape and lead to corneal transplantation. Hence, corneal  modeling plays an essential role in diagnosing and managing keratoconus to assess suitability of a subject for the treatment and prevent  improper refractive surgeries \cite{Maeda:1995fk}. Also, the great role of the reliable visualization tools in clinical practice should not be underestimated.

Modern techniques of design and fit of soft contact lenses can take into account features of the patient's eye, adapting the back surface of a lens to match the specific elevations of the cornea. These methods require again a detailed corneal topographic analysis of the anterior face of the cornea. 

With the introduction of the high-speed videokeratoscopy \cite{Nemeth:2002kx, Iskander:2005vn} in the study of the dynamics of corneal surface topography \cite{Zhu:2021ys} and tear film stability \cite{Iskander:2005zr}, the storage needs have become significant, motivating another  important application of corneal surface modeling: data compression \cite{Schneider:2009uq}.

The vast majority of modern corneal topographers collects the data (either elevation, curvature, mire displacement or others) in a finite and discrete set of points. Typically, these points present a quasi-structure; for instance, those devices based on the Placido rings technology provide elevations and curvatures at the discretized images of the mires, whose deformation is proportional to the complexity of the surface. In any case, the data is contaminated by the error, which stems from several sources: the natural device noise, measurement and digitalization errors, algorithm errors (like those converting the displacement in elevation), rounding errors and others. Hence, we face the problem of the parsimonious fit of the actual surface data contaminated by noise, with a minimum number of coefficients or parameters, for its clinical and technological applications.

The solutions to this problem are usually classified into the so-called zonal and modal methods. In the former group, the domain where the data are collected is subdivided in more elementary subdomains (e.g., triangles), and the surface is approximated in each subdomain with a relative independence from the other regions. The standard tool here are splines, in particular, the numerically stable B-splines \cite{Halstead:1995ky, Liu:2003fk, Ares2006}, widely used in the Computer-Aided Geometric Design.

In the modal methods of reconstruction the approximation is found as a (typically, linear) combination of functions from the given set or dictionary, defined by a number of parameters. In this sense, crucial decisions to make are the set of functions to use, the value of their parameters, and the number of functions needed to recover the relevant information without fitting (at least, in a large scale) the inevitable error (the so-called model selection problem). 

Among the advantages of the zonal methods is the flexibility and the accuracy of fit, but they lack the simplicity of the modal approach, which renders functional expressions valid across the whole domain, suitable for further calculations (such as ray tracing and others). Zonal techniques are also substantially more computer-intensive and encode the final shape in a larger amount of data.

A standard functional basis for the modal reconstruction, used commonly in ophthalmology to express ocular wavefront error are the Zernike polynomials \cite{Klyce2004}. The coefficients of their expansions have interpretation in terms of the basic aberrations such as defocus, astigmatism or coma, along with higher order aberrations such as trefoil and spherical aberrations. As a fitting routine, Zernike polynomials are not limited to analysis of wavefront error surfaces, but can be applied to other ocular surfaces as well, including the anterior corneal surface \cite{Smolek2003, Carvalho2005}. It has been suggested that Zernike analysis may be applicable in the development of corneal topography diagnostic tools, using the Zernike coefficients as inputs into corneal classification of neural networks \cite{Carvalho:2005ij, Smolek:1997gy}, replacing or supplementing the currently used corneal indices included with many topography devices. 

However, potential limitations in this approach have been reported in the literature \cite{Klyce2004, Smolek2005}. There is a growing concern that the Zernike fitting method itself may be inaccurate in abnormal conditions. Furthermore, it is very difficult to assess a priori how many terms are necessary to achieve acceptable accuracy in the Zernike reconstruction of any given corneal shape \cite{Iskander2001}. It is known \cite{Smolek2005} that limiting Zernike analysis to only a few orders may cause incorrect assessment of the severity of the more advanced stages of keratoconus \cite{Schwiegerling1989}. This information is particularly needed in the discriminant analysis of the decease markers, or when selecting the numerical inputs for neural network--based diagnostic software such as corneal classification and grading utilities for condition severity.

In this sense, several alternatives to the modal least-square fit with Zernike polynomials  have been recently suggested. Some of them intent  to combine the modal and zonal approaches in order to preserve the best of both worlds \cite{Espinosa:2010uq}  or to use non-linear methods \cite{Schneider:2009uq}. The idea of the possibility of getting the accuracy and flexibility of the zonal methods within the framework of the linear modal approach by means of localized radial basis functions has been expressed in \cite{Martinez-Finkelshtein:2009kx}, but without developing an actual implementation or procedure. In this paper we describe an adaptive and multi-scale working algorithm for the parsimonious fit of the surface data, based on residual iteration with knot insertion, that allows to adapt the number of functions used in the reconstruction to the conditions of each cornea. 

The residual iteration is well-known in many branches of mathematics; it is related for instance with the iterative refinement methods of solution of systems of linear equations. In the context of purely radial basis functions an adaptive greedy approximation algorithm using interpolation has been proposed in \cite{Schaback:2000fk}. The adaptive increase of the number of basis functions as a technique used to improve the quality of a given initial approximation is also standard, see for instance \cite[Ch.~21]{Fasshauer07} for a general discussion and references.

The method proposed here allows to build iteratively an approximation function as a linear combination of anisotropic gaussian basis functions, implementing also a dynamical selection of the parameters and the management of noise. It can be used to reconstruct altimetric data, corneal power maps, and others. Although it has been tuned up for the Placido ring based keratoscopes, with the data nodes located in almost concentring rings, the technique is actually applicable to any scattered data approximation. Part of this approach was announced in \cite{Martinez-Finkelshtein:2010fk}.

\section{The fitting procedure}

\subsection{The general setting}

The input data is a 3D cloud $(x_k,y_k,z_k)$, $k=1, \ldots ,N$, corresponding to either elevation or corneal power $z_k$  by a corneal topographer at the node $\PP_k$ of the anterior corneal surface with Cartesian coordinates $(x_k, y_k)$. We will discuss the case when $z_k$ corresponds to elevation. Taking into account the global shape of the cornea, a standard procedure is to ``flatten'' the data by fitting it with the best-fit sphere \cite{Ahn:2001wd} of the form:
$$ S(x,y) = z_0 + \sqrt{ R^2 - (x-x_0)^2 - (y-y_0)^2 } $$
where $R$ and $(x_0,y_0,z_0)$ are its radius and the Cartesian coordinates of its center, respectively. Although the common practice is to fit with the standard linear least squares, better results are obtained with a weighted least square fit, using $(1+\|P_k\|)^{-1}$ as the weight, in accordance with the typical error distribution \cite{Tang:2000vn}. 

As a result of the previous step, the residual errors $\varepsilon ^{(1)}_k  =   z_k - S(x_k, y_k)$ contain both the relevant information at different scales and noise. Our aim is to fit these residuals by a function $E(x,y)$ in such a way that an analytic expression for the corneal height is given by 
\begin{equation} \label{analyticExp}
\text{Cornea}(x,y)=S(x,y) + E(x,y),
\end{equation}
In this way, $S$ accounts for the global shape of the cornea, while $E$ captures the small irregularities in the surface. Function $E$ is given by a linear combination of $n$ functions from a given dictionary,
\begin{equation} \label{E}
E(x,y)=E_n(x,y)=\sum_{j=1}^n c_j\, h_j(x,y).
\end{equation}
In an ideal setting, $n$ depends on the actual data, and should be large enough to allow all relevant information from the elevation measured modeled by $E$, but not too large to prevent from overparametrizing the problem and fitting pure noise. In order to circumvent the difficulties of the Zernike polynomials mentioned above we use as basis functions the gaussians of the form
$$
h(x,y)=\exp\left( -\| \boldsymbol{P}- \QQ\|_A^2 \right), \quad \PP=(x,y)^T,
$$
where  the superscript $T$ denotes the matrix transpose, $\QQ=(Q_x, Q_y)^T$ is a certain point on the plane (``center''), and $A$ is a positive-definite matrix in $\R^{2 \times 2}$. For such a matrix the $A$-norm of a point (column vector) $\PP$ in $\R^2$ is defined as 
\begin{eqnarray*}
&\| \PP \|_A  =  \sqrt{\PP ^T A \PP} =  \sqrt{\alpha_x \, x^2 + \alpha_y \,  y^2   + 2\alpha_{xy} \,  x y}, \\
& \text{for } A=\begin{pmatrix}
\alpha_x & \alpha_{xy} \\
\alpha_{xy} & \alpha_y
\end{pmatrix} \text{ with $\alpha_x>0$ and $\alpha_x \alpha_y> \alpha_{xy}^2$}.
\end{eqnarray*}

In general, these are anisotropic radial basis functions, that boil down to standard radial basis gaussian functions (RBGF) when both eigenvalues of $A$ coincide (in other words, when $A$ is a positive multiple of the identity matrix $I_2$). 

One of the advantages of these functions is that they are quasi compactly supported: for $|x|>1.7308$,  $e^{-x^2}<0.05$, that is, achieves less than 5\% of its maximum height. 

Hence, we seek the expression of the form
\begin{eqnarray*}
\text{Cornea}(x,y)=S(x,y)  + \sum_{j=1}^n c_j\, \exp\left( -\| \PP-\QQ^{(j)}\|_{A_j}^2 \right),\quad \PP =(x,y)^T.
\end{eqnarray*}
Clearly, a fitting routine should allow for an adequate selection of all parameters, namely
\begin{itemize}
\item centers $\QQ^{(j)}$;
\item shape matrices (or shape parameters) $A_j$;
\item scaling factors $c_j$;
\item number of terms $n$ in the functional representation.
\end{itemize}

We propose an iterative algorithm of reconstruction, such that in each step we fit partially the residual error by one anisotropic RBGF (A-RBGF), and compute the new residuals, which will become the input for the next iteration (residual iteration with knot insertion). To preserve the maximum possible degrees of freedom, the centers, the shape parameters and the scaling factors will be chosen dynamically depending on the residual data in each step.

\subsection{Description of the iterative algorithm}

Let $E_{j-1}$ be already computed (we take $E_0\equiv 0$). The input data for the $j$'s iteration ($j=1, 2, \dots$) is the cloud $(x_k, y_k, \varepsilon ^{(j)}_k)$ of nodes $\PP_k=(x_k, y_k)^T$ and the corresponding residuals $\varepsilon ^{(j)}_k$, $k=1, 2, \dots, N$; recall that $\varepsilon ^{(1)}_k  =   z_k - S(x_k, y_k)$ are the residual errors after the weighted best sphere fit. We perform the following steps:

{\sc Step 1:} selection of the center $\QQ^{(j)}$.

The problem of the selection of a center of a radial basis function has been discussed in \cite{Jamshidi:2007fk}. There the center is chosen among the data nodes $\PP_k$ using the criterion of maximum cross-correlation. Another criterion uses the power function (see e.g.~\cite{Schaback:2000fk}, \cite{Fasshauer07}, \cite{De-Marchi:2005uq}). Both methods, although computationally demanding, can be implemented to perform Step 1. However, in our practice we found the much simpler criterion of maximal residual (strategy for so-called greedy approximation) to be as satisfactory, at a minimum cost; it correlates also with the geometry of the A-RBGF. Hence, we choose 
$$
\QQ^{(j)}=(x_{k_0},y_{k_0})^T \quad \text{where} \quad k_0 = \arg \max_k |\varepsilon_k^{(j)}|
$$
and denote
$$
m^{(j)}:=\varepsilon_{k_0}^{(j)}.
$$

{\sc Step 2:} dynamical filtering.

As it was mentioned before, the altimetric data obtained from measuring devices such as a keratographer are contaminated by noise. Although there are some indications about statistical distribution of these errors, the information is still very limited. In order to cope with this problem we need to filter out those data that clearly correspond to the measurement error and thus spoil the quality of the reconstruction. We can do it in advance, before starting the fitting procedure, like in \cite{Alkhaldi:2009fk}. But we have chosen a simpler alternative, that yields satisfactory results: once the center $\QQ^{(j)}$ has been selected, we check the number, $\ell_k$, of nodes $P_k$ lying in the largest disk, centered at $\QQ^{(j)}$ and containing only nodes with the residues of the same sign as $m^{(j)}$. If  $\ell_k <20$, we consider $\QQ^{(j)}$ an outlier and exclude it from consideration at this iteration. This can be done by simply setting $\varepsilon_{k_0}^{(j)}=0$, after which we return to Step 1. Otherwise, we proceed to the next step.

Obviously, this step can be ignored if we know that the error is negligible.

{\sc Step 3:} selection of the shape parameters.
 
We determine first the influence nodes $\mathcal P_j(s)$, defined as the maximal set of at most $ s$ nodes $\PP_k$ closest to $\QQ^{(j)}$  with residues of the same sign than $m^{(j)}$. Observe that $\mathcal P_j(s_1)\subset \mathcal P_j(s_2)$ if $s_1 \leq s_2$. It is convenient to parallelize the subsequent computations for several values of $s$: we have performed experiments using the vector of values $s=[s_{\min}, 100, 150, 200, 300]$, where $s_{\min} =\min ( \ell_k, 50)$, with $\ell_k$ defined in Step 2. 
 
The interpolating conditions
$$
\varepsilon_{k_0}^{(j)} h_j(x_k, y_k)= \varepsilon_{k}^{(j)}, \quad k \in \mathcal P_j(s) ,
$$
are equivalent to the overdetermined linear system 
\begin{equation} \label{system}
\begin{split}
\alpha_x \, (x_k-x_{k_0})^2  + 2\alpha_{xy} \,  (x_k-x_{k_0})(y_k-y_{k_0})  \\+ \alpha_y \,  (y_k-y_{k_0})^2   
=\log \left( \frac{\varepsilon_{k_0}^{(j)}}{\varepsilon_{k}^{(j)}} \right), \quad k \in \mathcal P_j(s)
\end{split}
\end{equation}
in the 3 unknown entries of the shape matrix
$$
A_j=\begin{pmatrix}
\alpha_x & \alpha_{xy} \\
\alpha_{xy} & \alpha_y
\end{pmatrix}.
$$
We solve this system in the sense of weighted linear least squares (WLS), where the $k$-th equation is multiplied by the weight $\eta_k:=(1+\|\PP_k-\QQ^{(j)}\|^2)^{-1}$ in order to account for the bigger influence of the neighboring nodes on $A_j$. This solution is obtained by standard methods, using either the $QR$ factorization of the collocation matrix corresponding to \eqref{system} or its singular value decomposition, see e.~g.~\cite{trefethen/bau:1997}.

Observe that due to the selection of the active center $\QQ^{(j)}$,
$$
t_k:=\log \left( \frac{\varepsilon_{k_0}^{(j)}}{\varepsilon_{k}^{(j)}} \right)\geq 0, \quad k \in \mathcal P_j(s).
$$
However, this condition does not guarantee that the solution $A_j$ of  \eqref{system} in the sense of the WLS described above will be positive definite. This can typically fail in the periphery of the convex hull of the nodes, where the lack of data in some direction might yield non-positive definite $A_j$. Although the corresponding function $h_j$ might fit the data correctly locally, it is not valid globally due to its exponential increase in the direction of the eigenvector of a negative eigenvalue of $A_j$.

In order to overcome this problem we examine the solution $A_j$ of \eqref{system}: if it is not positive definite, we interpret that there is a lack of data in a neighborhood of $\QQ^{(j)}$ and force $h_j$ to be an isomorphic (a bona fide) radial basis function: $A_j=\alpha I_2$. In this way, \eqref{system} is reduced to
\begin{equation} \label{systemRBF}
\alpha  \, \left \| \PP_k-\QQ^{(j)}\right\|^2  =t_k, \quad k \in \mathcal P_j(s), 
\end{equation}
whose solution in the sense of the WLS is
$$
\alpha  = \sum_{k \in \mathcal P_j(s)}  \theta_k\,  t_k,
$$
with 
\begin{equation*}\label{formulaA} 
\theta_k  =\frac{  \eta_k^2 \left \| \PP_k-\QQ^{(j)}\right\|^2 }{\sum_{t \in \mathcal P_j(s)}  \eta_t^2 \left \| \PP_t-\QQ^{(j)}\right\|^4 }
\end{equation*}
 and $\eta_k=\left(1 +\left \| \PP_k-\QQ^{(j)}\right\|^2 \right)^{-1}$.
Observe that in this case $\alpha$ is positive by construction, and we  define
$$
h_j(x,y)=\exp\left( -\alpha\,  \| \PP-\QQ^{(j)}\|^2 \right),\quad \PP =(x,y)^T.
$$

{\sc Step 4:} selection of the scaling factor.

We can calculate the coefficient $c_j$ from
$$
c_j\,  h_j(x_k, y_k)= \varepsilon_{k}^{(j)}, \quad k \in \mathcal P_j(s) ,
$$
using the WLS with the same weights $\eta_k$:
$$
c_j = \sum_{k \in \mathcal P_j(s)}  \gamma_k\,  \varepsilon_{k}^{(j)},
$$
with
$$
\gamma_k = \frac{  \eta_k^2 \, h_j(x_k,y_k) }{\sum_{t \in \mathcal P_j(s)}  \eta_t^2 \, h_j(x_k,y_k)^2  }.
$$

It should be noted however that numerical experiments show that in many cases the much simpler interpolation condition $ c_k=m^{(j)}$ yields comparable results.

{\sc Step 4:} computation of the new residuals.

With the values of $c_j$ and $A_j$ just computed we update
\begin{align*}
\varepsilon_k^{(j+1)}= \varepsilon_k^{(j)}  - c_j\,  h_j(x_k,y_{k}) .
\end{align*}
As it was mentioned before, all the computations have been performed in parallel for different values of $s$ (typically, from 3 to 5 values between 50 and 300), and hence, different nested sets of influence nodes $\mathcal P_j(s)$. We now keep the value of $s$ (and the corresponding values of $c_j$ and $A_j$) that yields the smallest norm of the residue vector $(\varepsilon_k^{(j+1)})$, and discard the other values. As a result, we find the new approximation, $  E_j = E_{j-1}+ c_j\, h_j$.

As a final step, we check the stopping criterium that will be discussed below. If this is not satisfied, we increment the iteration counter $j$ in 1 and return to Step 1.

\subsection{Stopping criteria} \label{subs:stopping}

In theory, the algorithm run indefinitely yields an interpolating function, and in consequence, a zero residue vector. In the real life situation of the data contaminated by errors, a very important problem is that of the model order selection: we want to capture all the relevant information without over-parametrizing the model and without fitting the noise. Many solutions to this problem are described in the literature. For instance, the choice of the number of Zernike polynomials for the modal reconstruction of the altimetric data has been discussed in \cite{Iskander2002, Iskander2008}.  

The statistical methods of selection of the appropriate number $n$ in \eqref{E} usually make assumptions about the noise. 
However, a priori information about the measurement error bounds or measurement error distribution is limited. According to \cite{Tang:2000vn}, the errors cannot be assumed i.i.d.\ random variables, although the assumption that they are normally distributed (with the variance proportional to the square of the distance of the node to the center) is apparently reasonable. They are also computationally intensive, \cite{Iskander2008, Iskander:2008kx}. 

Less demanding methods use information theory criteria, such as the Akaike Information Criterion (AIC), or the Efficient Detection Criterion (EDC) \cite{Rao:1989uq}, which studies the evolution of 
$$
EDC_j(p)=N \log\left(MSE_j\right) + j (N \log N)^p, \quad 0<p<1,
$$
with  
\begin{equation} \label{MSE}
MSE_j=\frac{\rho^2}{N}\, \sum_{k=1}^N \left(\varepsilon_k^{(j)}\right)^2, \quad \rho=\max_k \|P_k\|.
\end{equation}
The value of $p$ is usually tuned up experimentally.

However, we can gain information analyzing directly the behavior of the normalized errors $MSE_j$ defined in \eqref{MSE}. Typically, these errors start decaying with an exponential rate and average order greater than $1$. After a number of iterations we observe a stabilization in this rate of decay that becomes linear; this typically happens when values of $MSE_j$ are between $10^{-3}$ and $10^{-4}$ $\mu$m$^2$. Based on this experience we have used successfully the following stopping criterion: we allow the algorithm run for up to 50 iterations (this takes less than 2 seconds to complete) and calculate the weighted slopes
$$
\delta MSE_j = \frac{\log(MSE_{j+1}/MSE_j)}{j}.
$$ 
The sequence $\delta MSE_j $, although oscillatory, is negative and tends to zero, so we find the last iteration $1\leq J_1 \leq 50$ when $\delta MSE_j \geq -0.02$. If $ MSE_{J_1} < 10^{-3}$ $\mu$m$^2$, we fix $J_1+1$ as the stopping iteration. Otherwise we seek for the last iteration $1\leq J_2 \leq 50$ when $\delta MSE_j \geq -0.01$, and stop the algorithm at the $(J_2+1)$-th iteration.

\section{Experimental results}

In this section, we present a comparison of the numerical results obtained with our method applied both to simulated and real cornea surfaces. All the procedures were implemented in Matlab 7 and run on standard platforms (Windows PC and Mac with average configuration). The altimetric and curvature data from in-vivo corneas used for experiments described below were collected by the CSO topography system (CSO, Firenze, Italy), which in ideal conditions digitizes up to 24 rings with 256 equally distributed points on each mire. 

Since the procedure is meant for real-time reconstruction of the corneal data, any extremely computer-intensive methods should be discarded. However, in our Matlab implementation the execution time was always below 2 seconds, so this becomes less and less of a concern with the progress of the software optimization and computing power.

\begin{figure*}   
\begin{tabular}{cc}  
\includegraphics[scale=0.45]{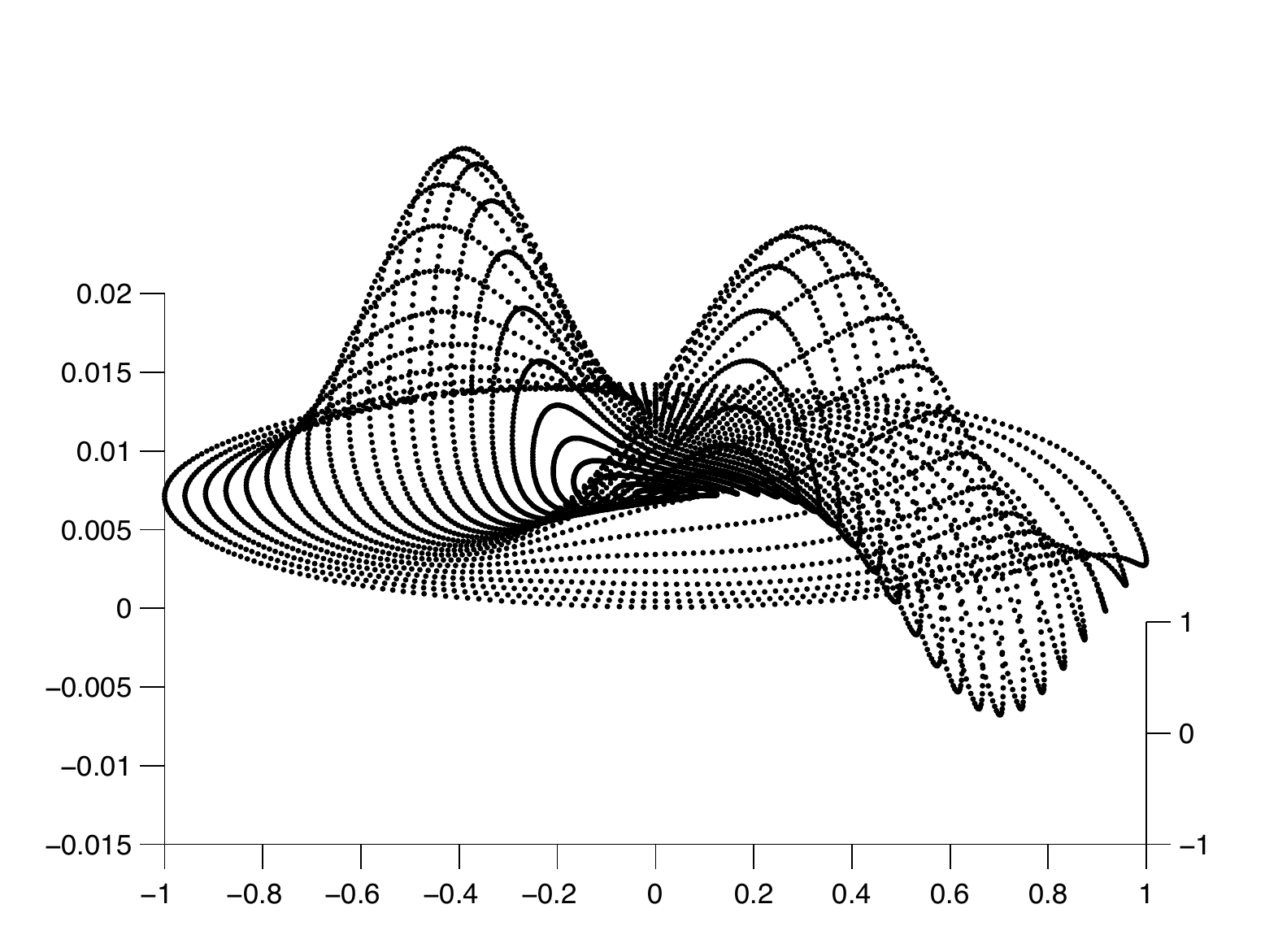} & \includegraphics[scale=0.45]{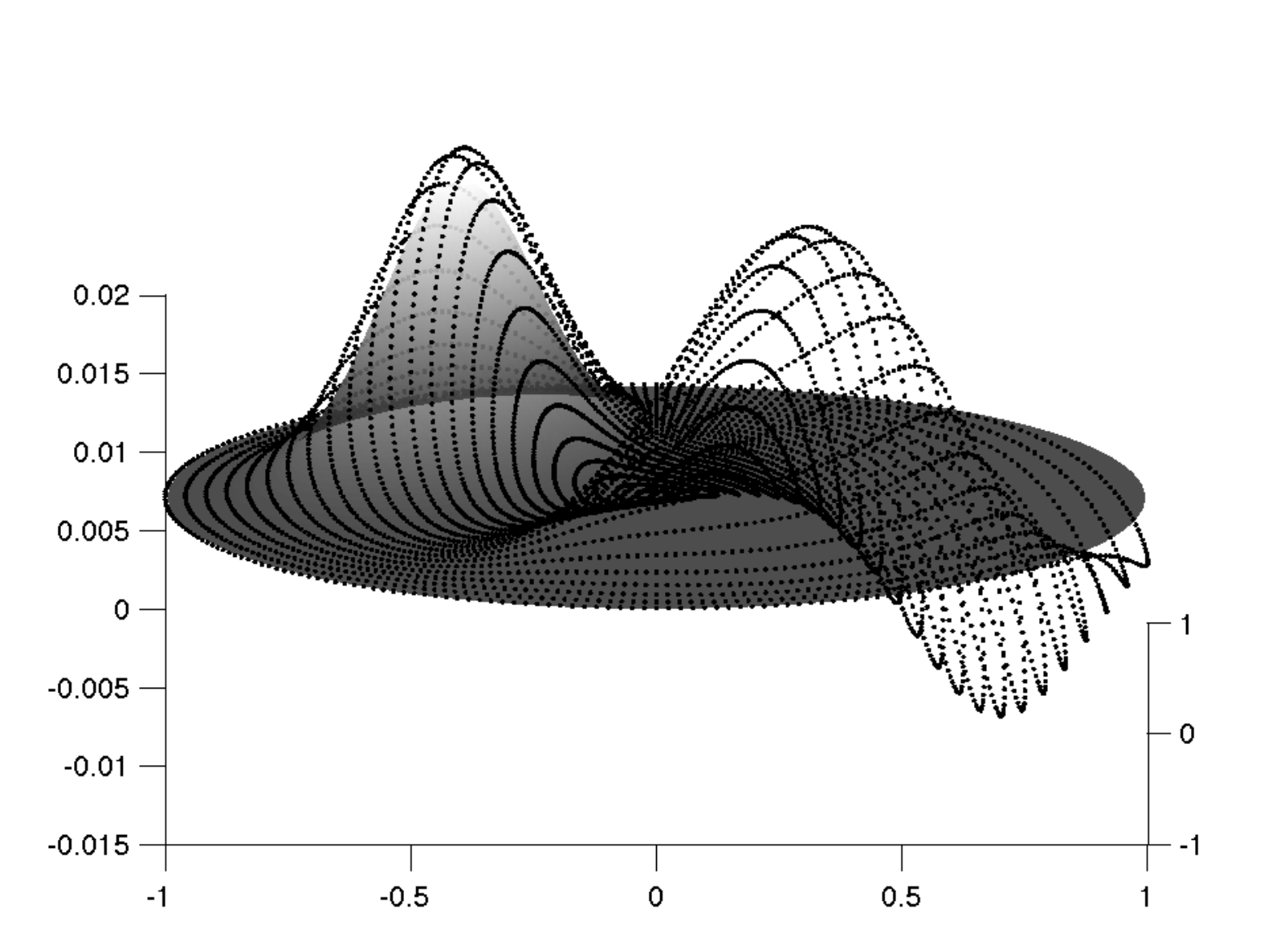} \\
\includegraphics[scale=0.45]{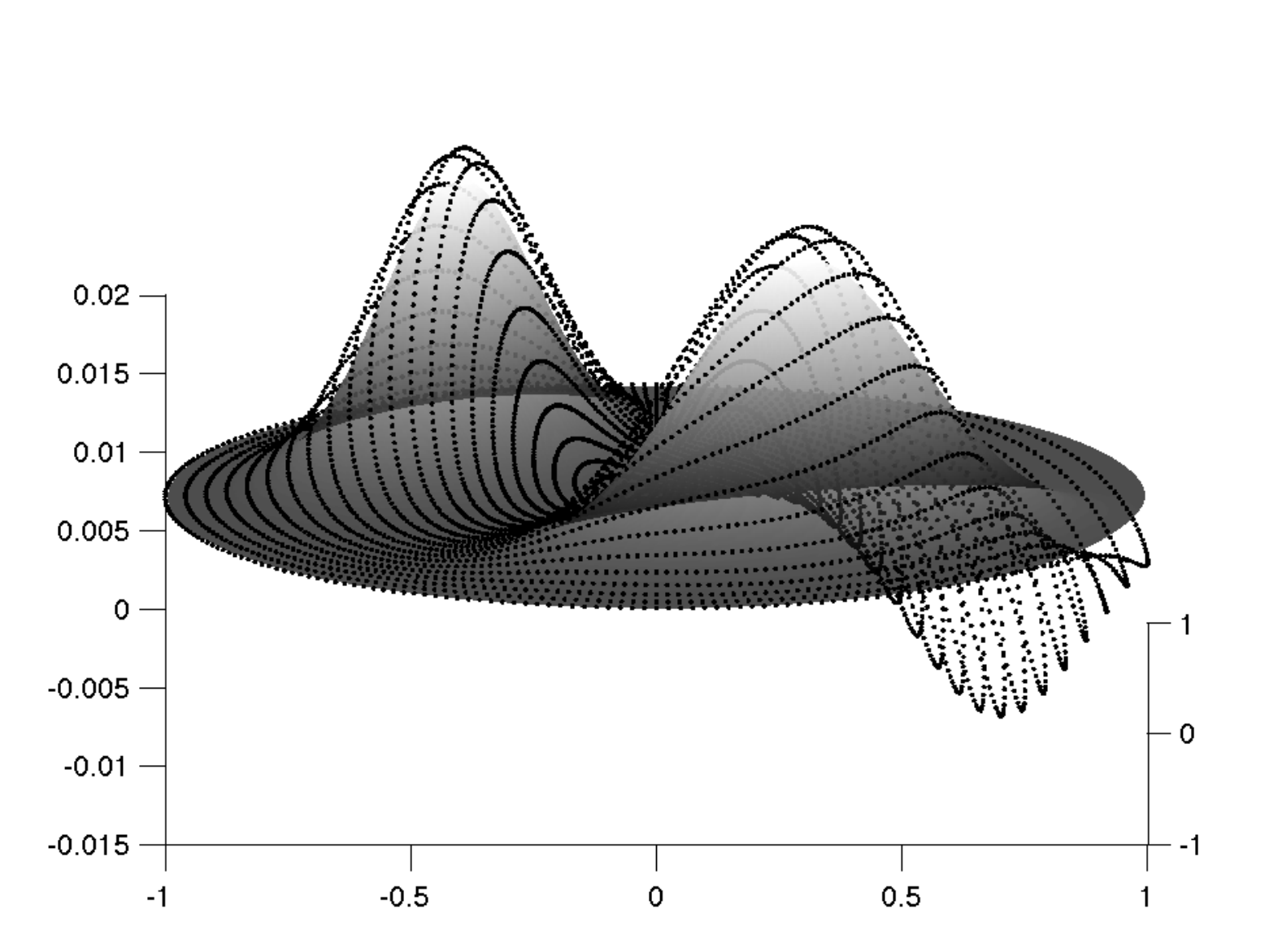} & \includegraphics[scale=0.45]{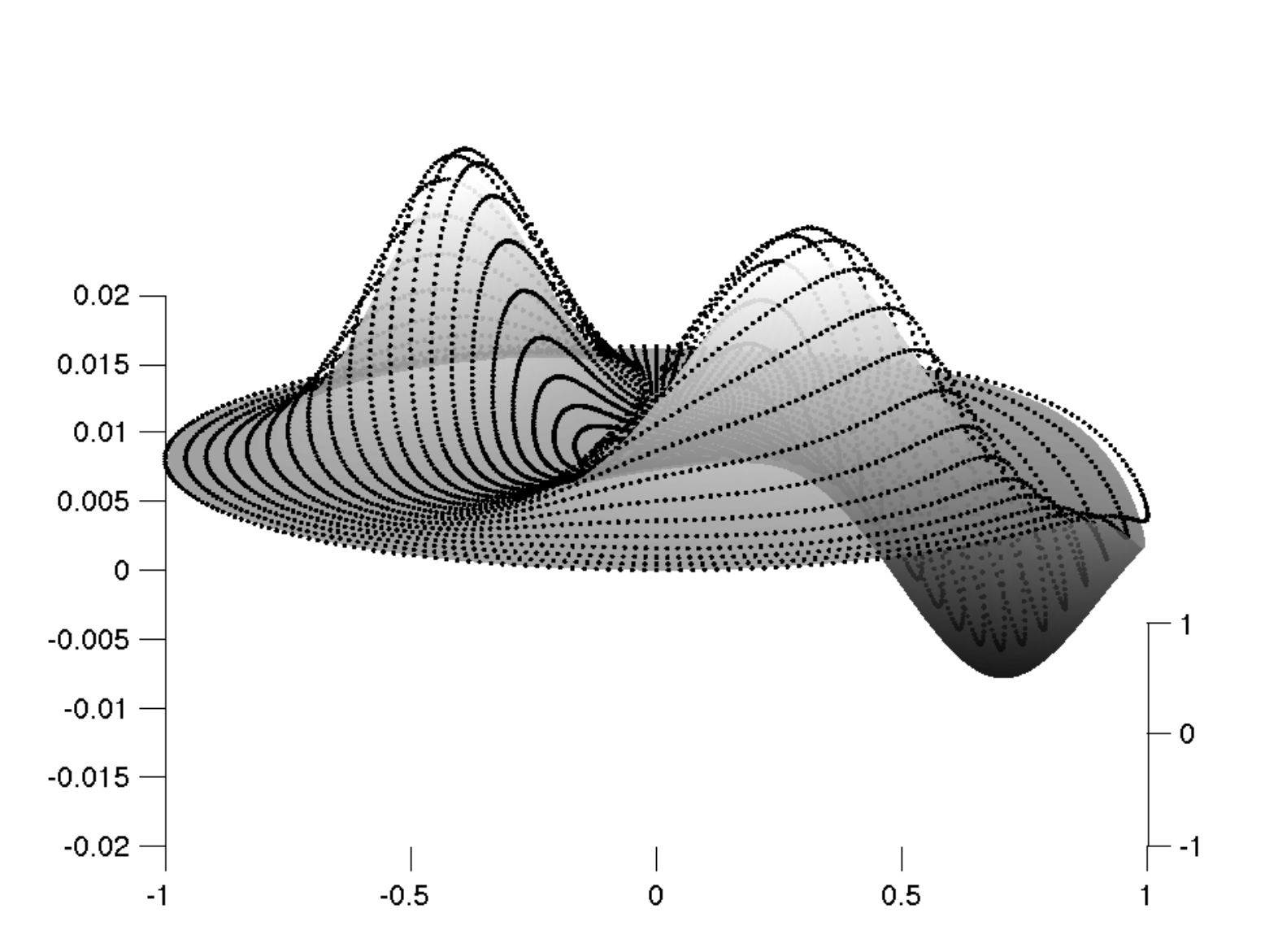}
\end{tabular} 
\caption{\label{figs:3peaks} Example of the performance of the algorithm in the simplest case of a surface comprised of three gaussians (upper left): the first iteration (upper right) matches the highest peak, while the next two capture the rest of the features of the surface.}
\end{figure*}

We demonstrate first the power of the proposed methodology using some elementary surface models, starting with the simplest example: a surface given by a linear combination of three exponentials,
$$
\text{Cornea}(x,y)=\sum_{j=1}^3 c_j \exp\left( - \| \PP - \QQ_j \|_{A_j}^2 \right), \quad \PP=(x,y)^T,
$$
with $\QQ_1=(0.3, -0.4)^T$, $\QQ_2= (0.7, 0.1)^T$, $\QQ_3=( -0.4, 0.3)^T$, $c_1 = 0.02$, $c_2= -0.015$, $c_3= 0.02$, and
$$
A_1=\begin{pmatrix}
10 & -7 \\
- 7 & 20
\end{pmatrix}, \quad A_2=\begin{pmatrix}
15 & 3.5 \\
3.5 & 10
\end{pmatrix},  \quad A_3=\begin{pmatrix}
20 & -6 \\
- 6 & 12
\end{pmatrix}.
$$ 
The surface is obviously exaggerated with respect to the typical residue of the standard cornea, but this is made with illustrative purpose. The result of the first three iterations of the algorithm are shown in Figure~\ref{figs:3peaks}.

\begin{figure*}   
\begin{tabular}{cc}  
\includegraphics[scale=0.5]{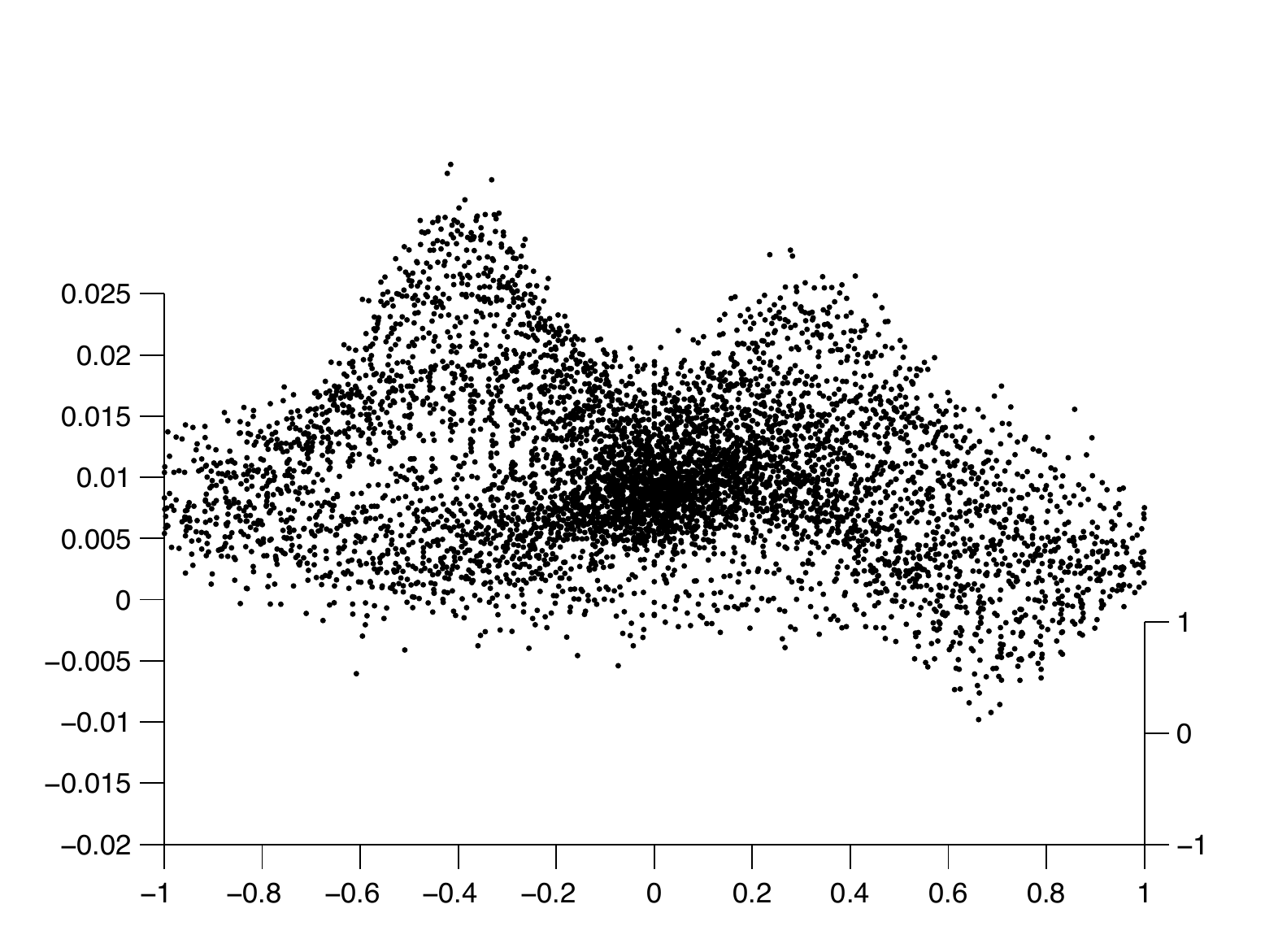} 
&  \includegraphics[scale=0.4]{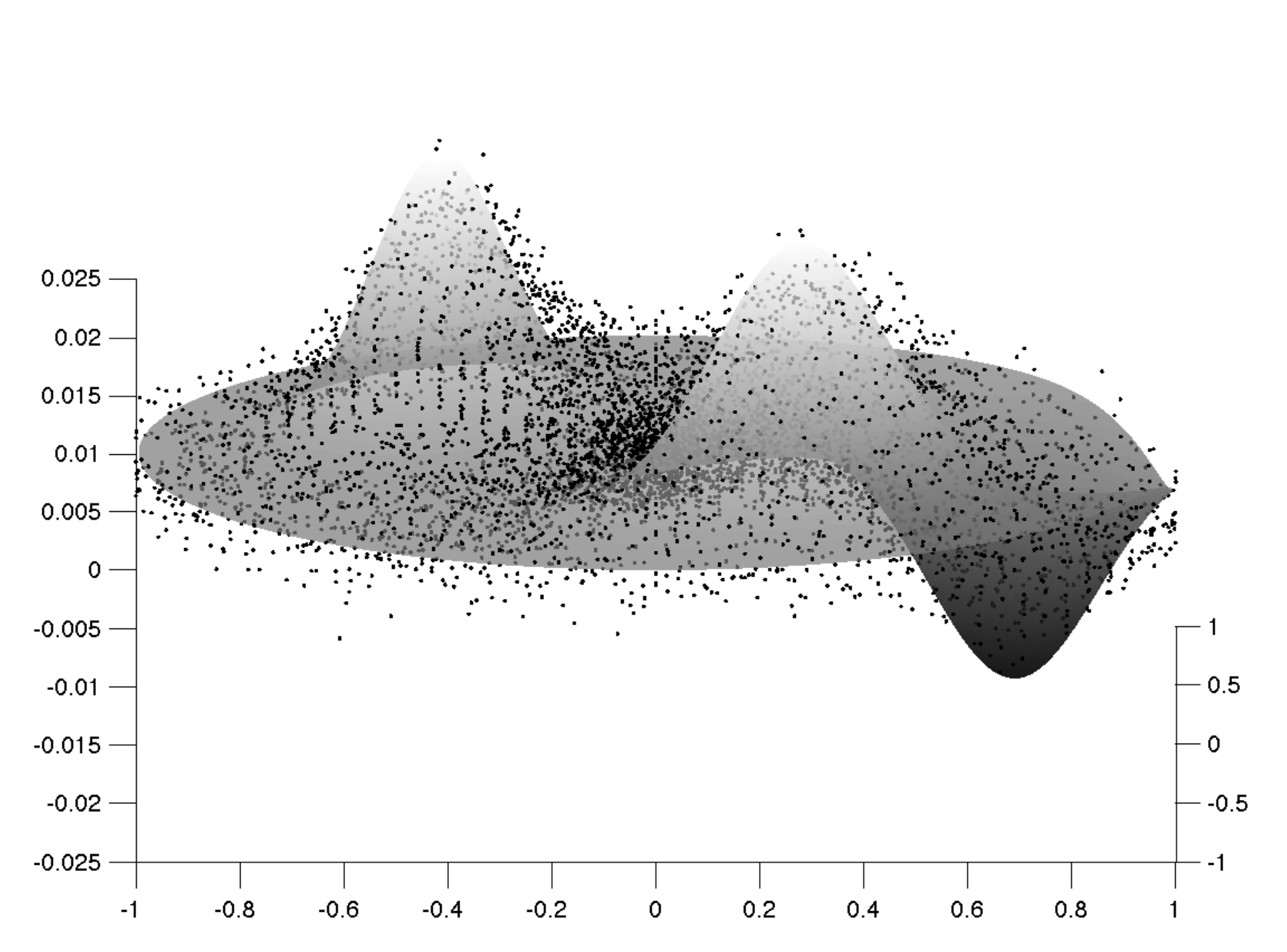}
\end{tabular} 
\caption{\label{figs:3peaksNoise} Performance of the algorithm with the surface analyzed in Figure~\ref{figs:3peaks} contaminated by white noise of 10\% of the maximum height: the data (left) and the result of the first three iterations (right).}
\end{figure*}

The measurement white noise (zero-mean and Gaussian noise process) of variance of 10\% of the maximum height was added to the surface. In most optical applications this would correspond to a very high level of the measurement noise. It should be noted that the knowledge of the distribution of the measurement noise is not necessary for our algorithm. As a result, the three centers of the Gaussians were correctly estimated (see Figure~\ref{figs:3peaksNoise}).

\begin{figure*}   
\begin{tabular}{cc}  
\includegraphics[scale=0.45]{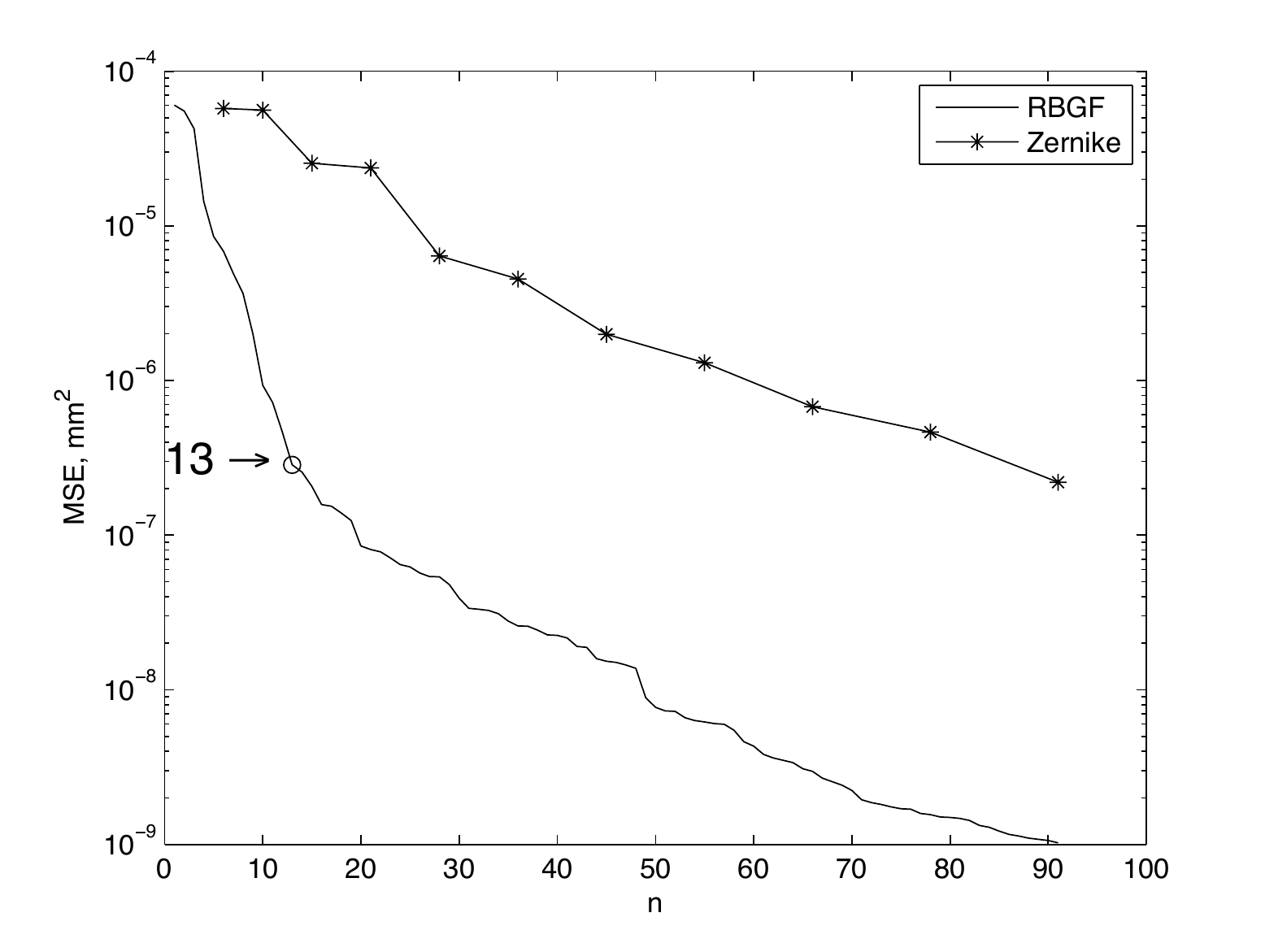} & \includegraphics[scale= 0.45]{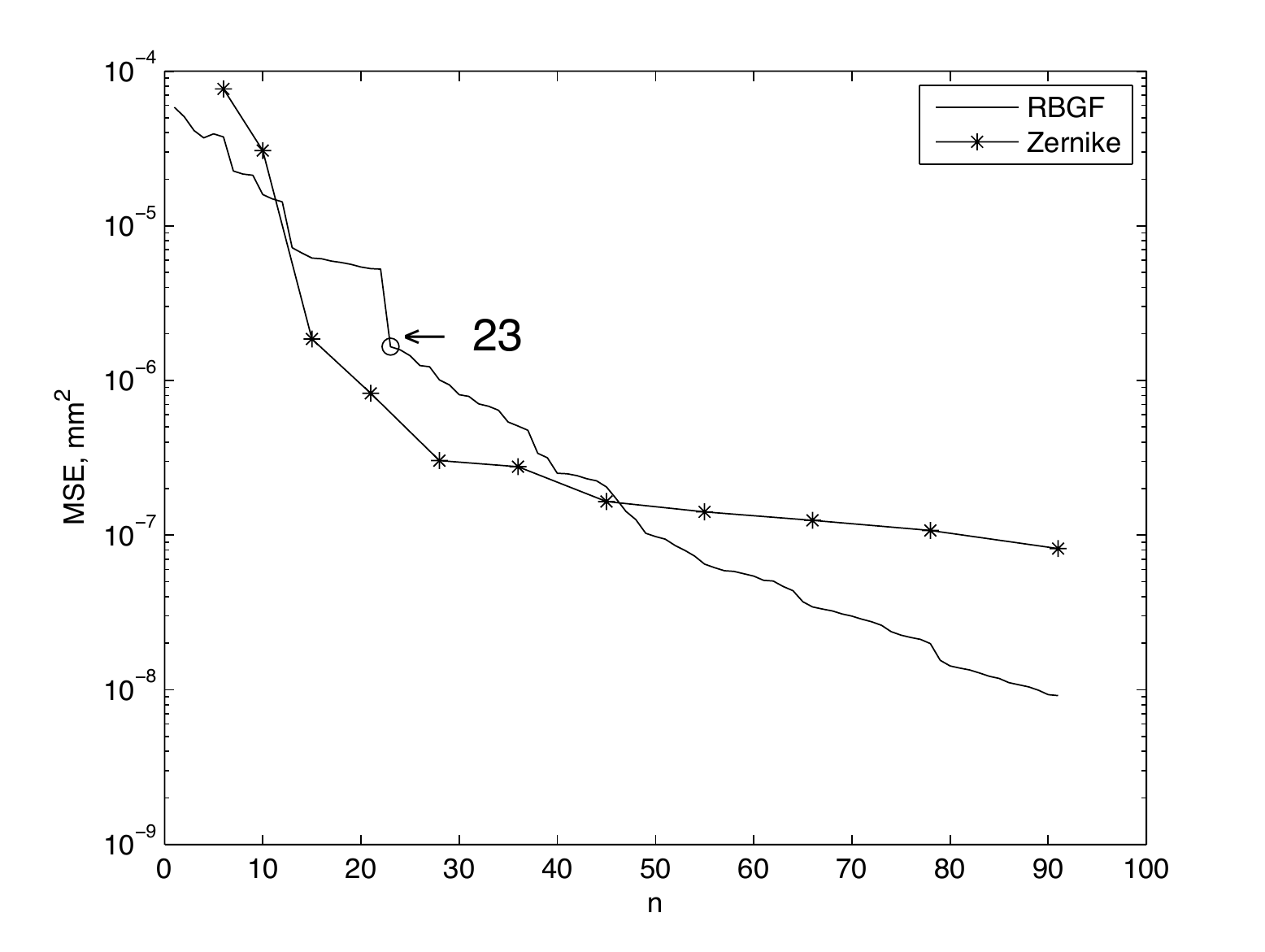} \\ \includegraphics[scale= 0.45]{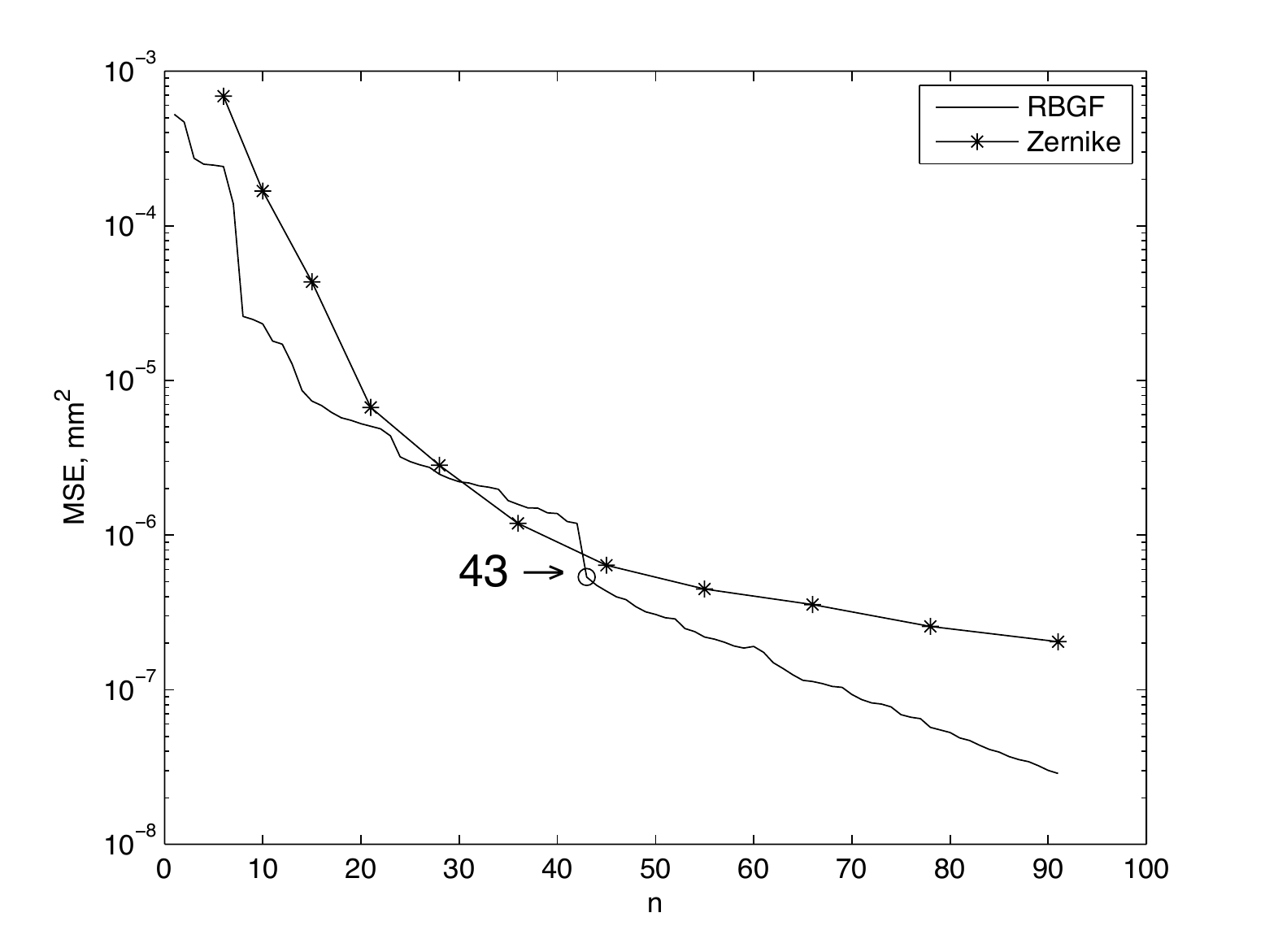} &
\includegraphics[scale= 0.45]{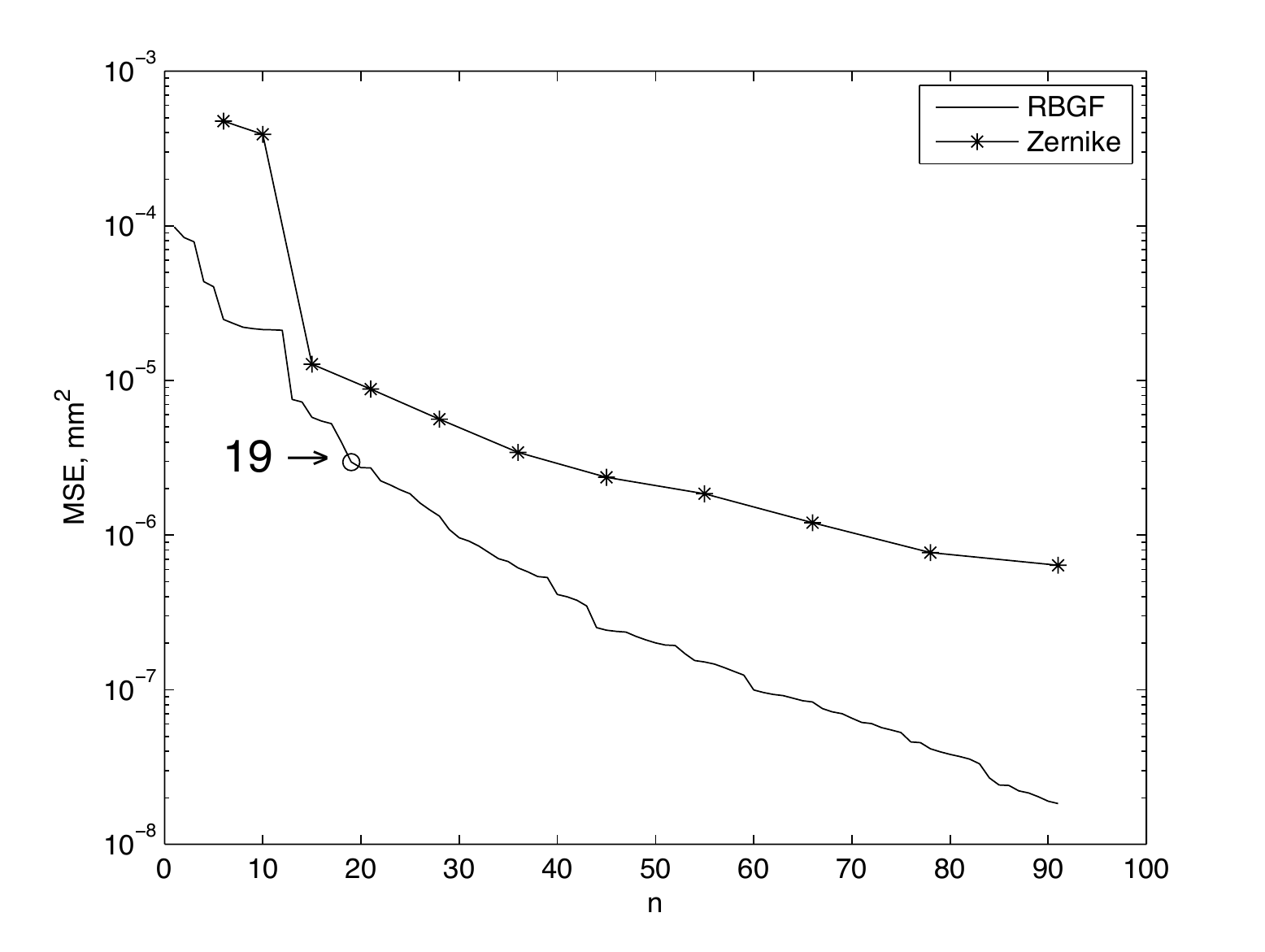}  
\end{tabular} 
\caption{\label{figs:Zernike} Comparison of the $MSE_j$ for altimetric data reconstruction by Zernike polynomials and by the A-RBGF algorithm. Upper left: synthetic cornea made of a linear combination of 10 gaussian functions. Upper right: normal cornea. Lower row: keratoconic corneas. The arrows indicate the stopping time of the algorithm according to the criterion described in Subsection \ref{subs:stopping}. The horizontal axis ($n$) indicates the number of functions used.}
\end{figure*}


The global error, measured by $MSE_j$, tends to decrease monotonically with the exponential rate. A typical behavior for the reconstruction of the altimetric data for both synthetic and real corneas can be observed in Figure~\ref{figs:Zernike}. In particular, in all cases we observe the feature mentioned above of the clear transition from an over-exponential to linear rate of decay, that is used as the stopping criteria. For comparison, we have reconstructed the same data with the Zernike polynomials using the linear least squares, which is the standard procedure in the clinical practice, implemented in most topographers. The $MSE_j$ for the Zernike reconstruction is plotted in the same Cartesian coordinate system, where $j$ indicates the total number of Zernike polynomials employed. Recall that $j$ varies from 1 to $ (m+1)(m+2)/2$, where $m$ is the maximal radial order used.

We observe two curious features of these plots. First, the Zernike polynomials easily capture the global shape of the reconstructed surface, which is expressed in a typical fast decay of the corresponding error. However, smaller details on the surface (such as areas of localized steepening) are much less suited for this tool. It explains the clear saturation observed in the Zernike error behavior after a few (typically, between 21 and 36 polynomials, corresponding to $5 \leq m \leq 7$. This is not the case of the reconstruction with A-RBGF, whose multi-scale and adaptive character allows to adjust the parameters adequately in each iteration.

On the other hand, numerical experiments show another interesting phenomenon related with the stopping criterion we use: in a majority of situations the stopping time corresponds to a $j$ for which $MSE_j$ is approximately equal for A-RBGF and Zernike polynomials.

However illuminating these graphs are, the global error is not the best way to compare both reconstruction approaches. Recall that the modal method with Zernike polynomials is suited precisely to achieve a maximum reduction of the $MSE_j$ for each $j$, while the iterative algorithm presented here has a totally different goal: locating the most salient feature of the residual surface and incorporating it into the analytic expression \eqref{analyticExp}. 

\begin{figure*}   
\begin{tabular}{cc}  
\includegraphics[scale=0.4]{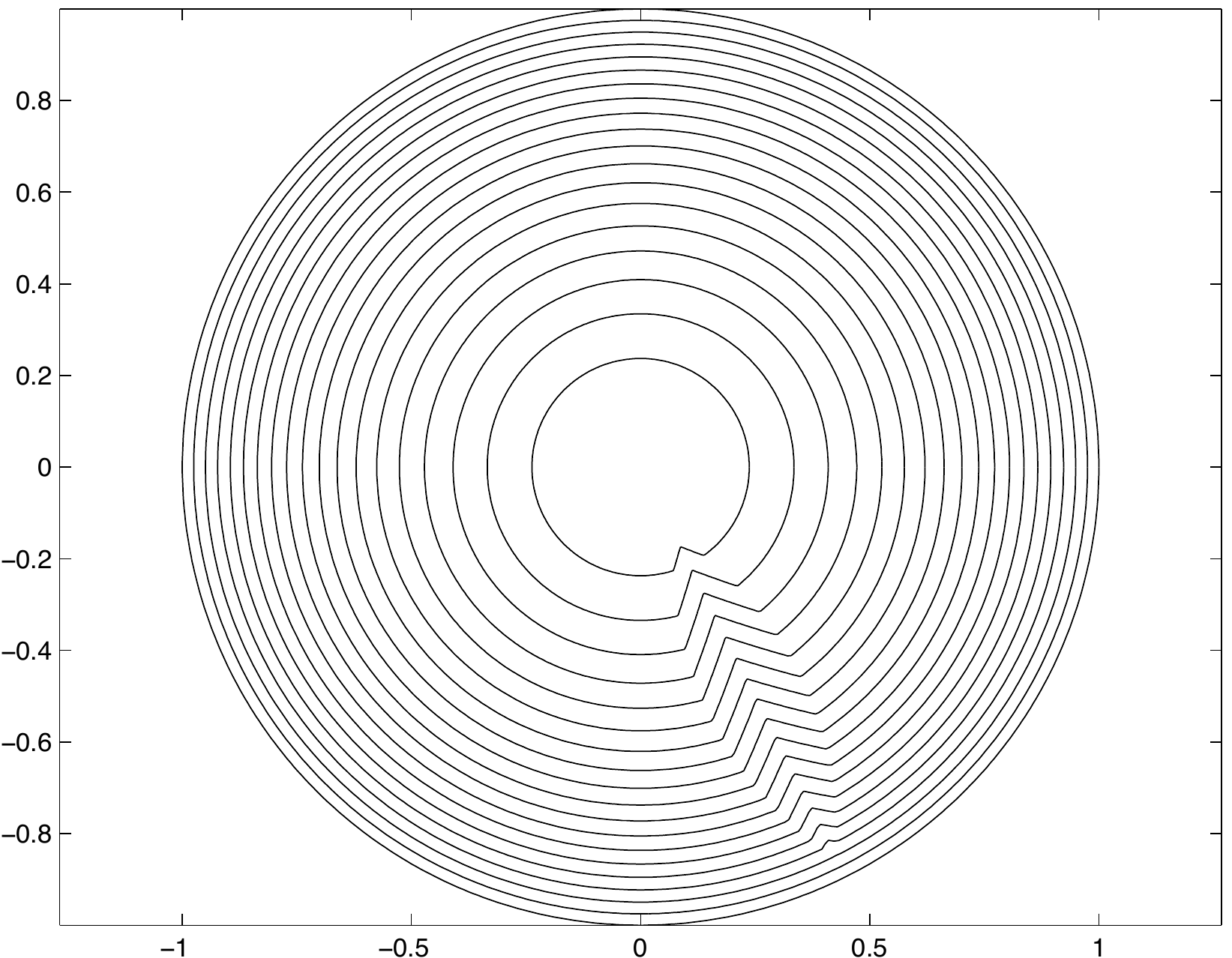} & \includegraphics[scale= 0.4]{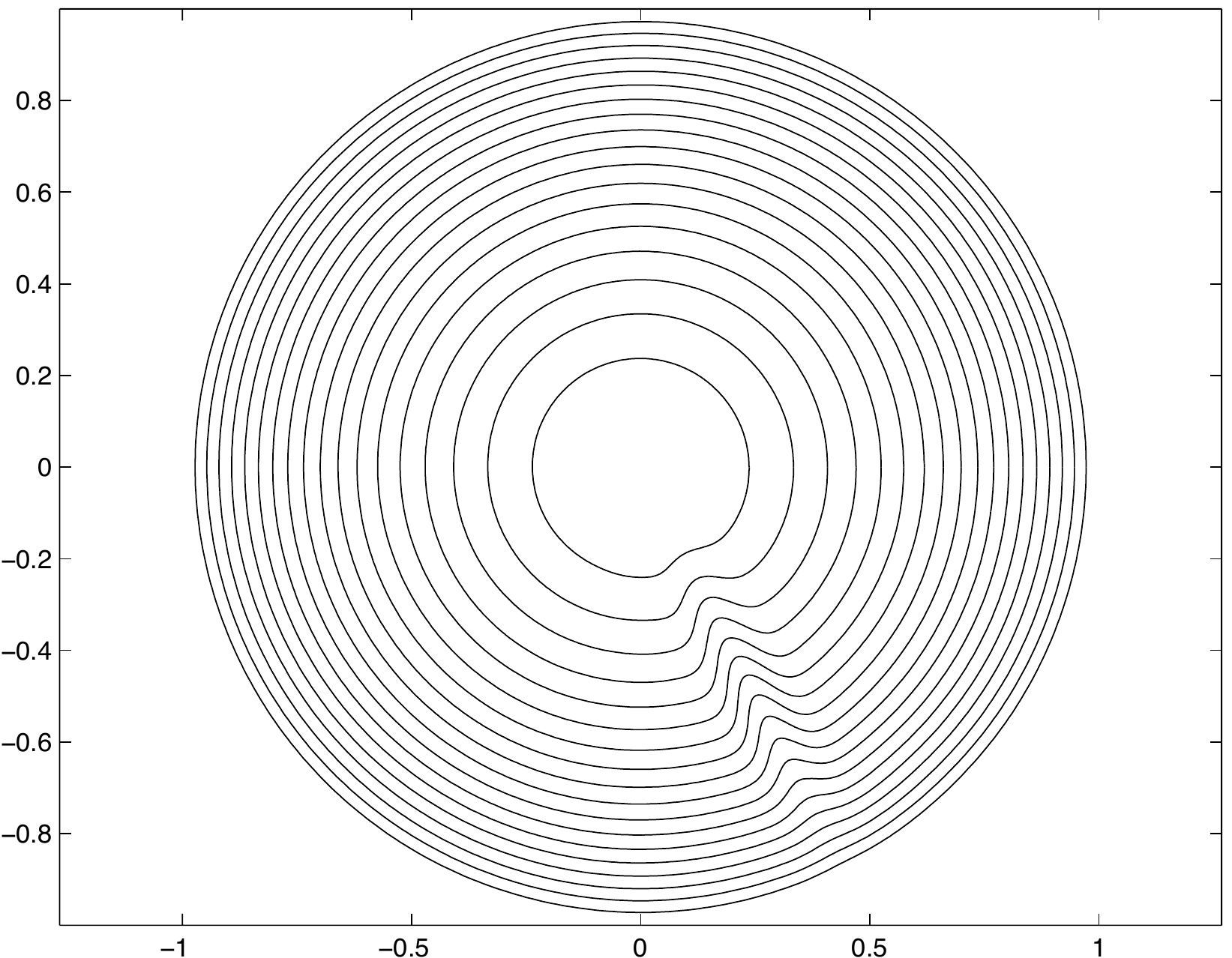} \\ \includegraphics[scale= 0.4]{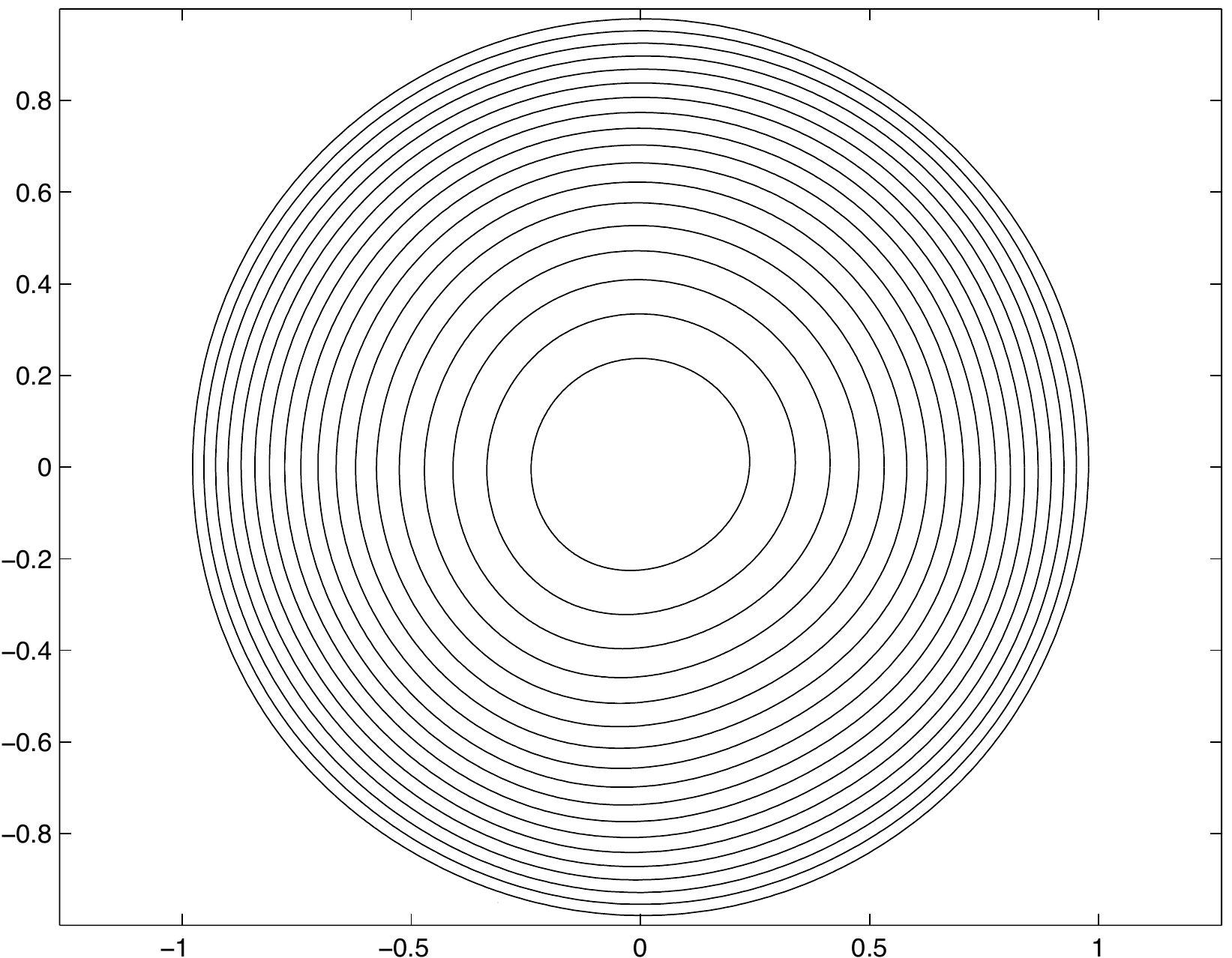} &
\includegraphics[scale= 0.4]{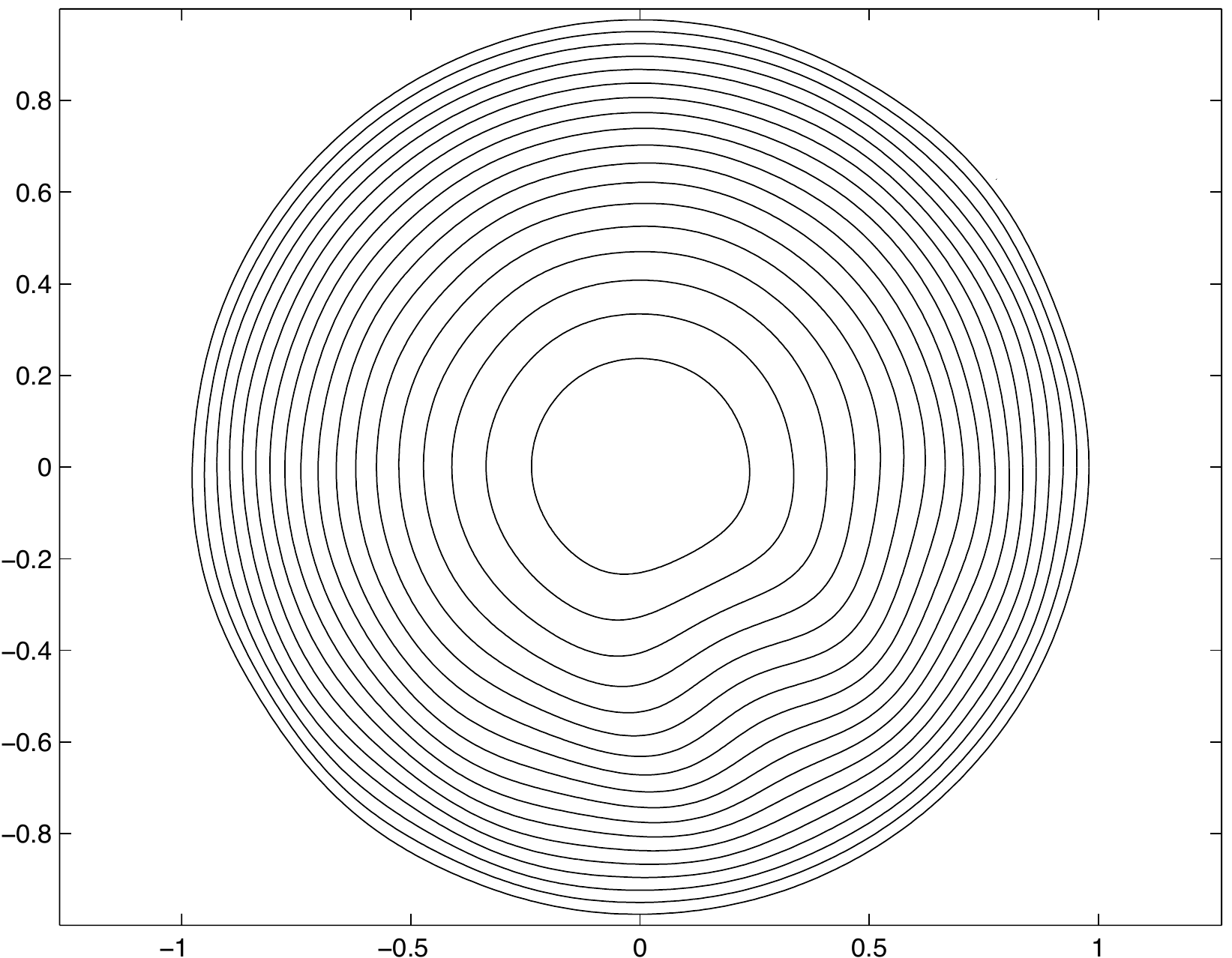}  
\end{tabular} 
\caption{\label{figs:Z} Comparison of the reconstruction of a synthetic ``scar'' (upper left) with 20 iterations of the adaptive algorithm (upper right), and 36 (lower left) and 136 (lower right) Zernike polynomials.}
\end{figure*}

This can be illustrated by fitting a synthetic ``cornea'' having a simulated scar, used already in \cite{Martinez-Finkelshtein:2009kx}. Its contour plot is depicted in Figure~\ref{figs:Z}, upper left. The upper right plot shows the contour plot of the surface reconstructed with the adaptive procedure described here using as few as 20 functions. The other two contour plots correspond to the same surface reconstructed with 36 (order $\leq 7$) and 136 (order $\leq 15$) Zernike polynomials.

\begin{figure*}   
\begin{tabular}{cc}  
\includegraphics[scale=0.48]{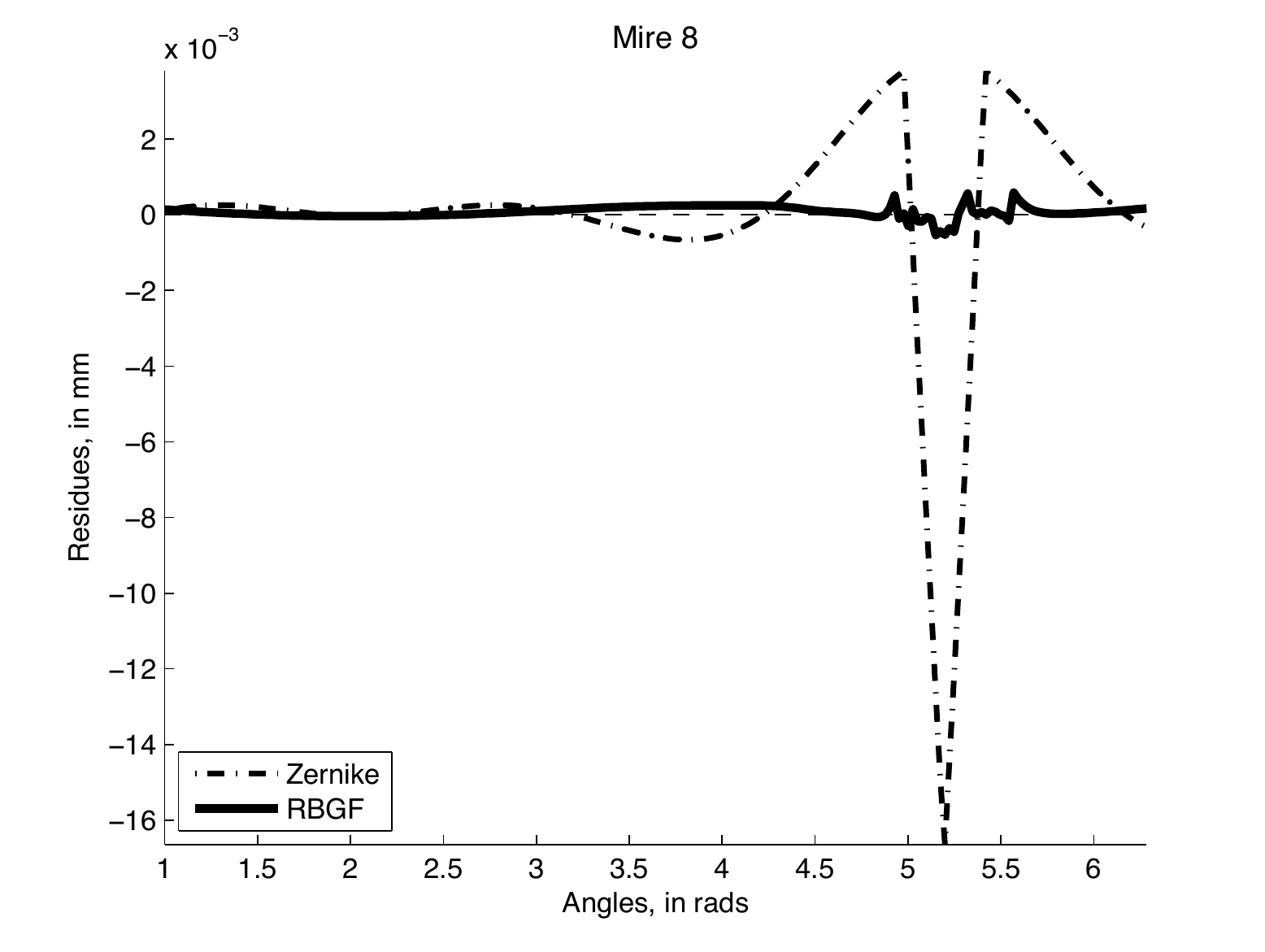}  & \includegraphics[scale=0.48]{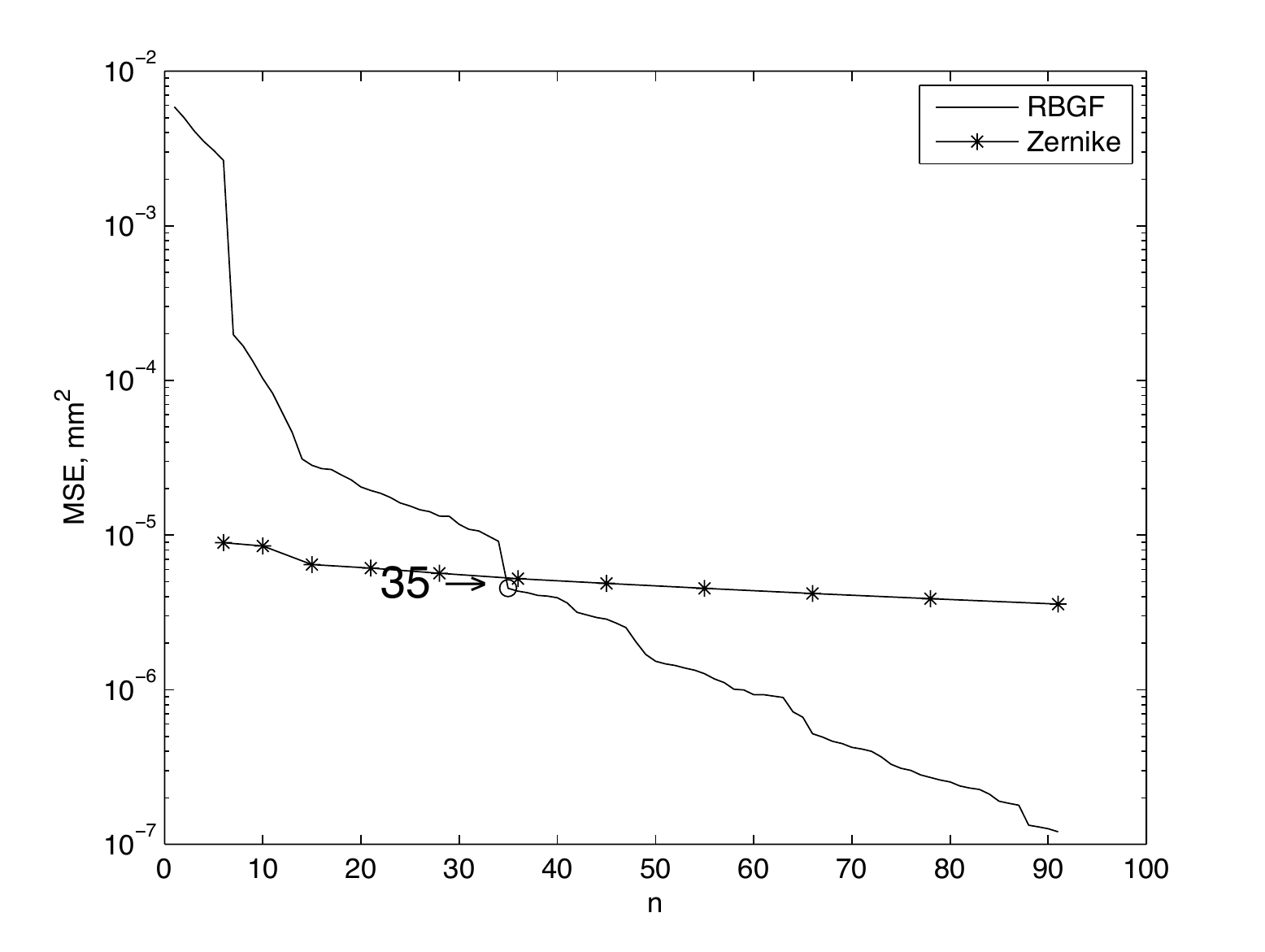}
\end{tabular}
\caption{\label{figs:mire} Left: residual errors along the mire 8 for the synthetic ``scar'' reconstructed with 20 iterations of the adaptive algorithm (bold line) and up to 36 Zernike polynomials (dotted line). Right:  $MSE_j$ for the ``scar'' reconstruction by Zernike polynomials and by the A-RBGF algorithm.}
\end{figure*}

The situation becomes even more apparent if we compare the residual errors along the mire 8 for both methods (Figure~\ref{figs:mire}, left): while the Zernike polynomials work perfectly on smooth portions of the surface, they find difficulties in adapting to fast varying shapes, where the A-RBGF algorithm uses its multi-scale and adaptive character to fit the surface almost perfectly after a few iterations. It takes a really big time for the $MSE$ errors of Zernike polynomials to progress, as illustrated in Figure~\ref{figs:mire}, right.

\begin{figure}   \centering
\begin{tabular}{cc}  
\includegraphics[scale=0.43]{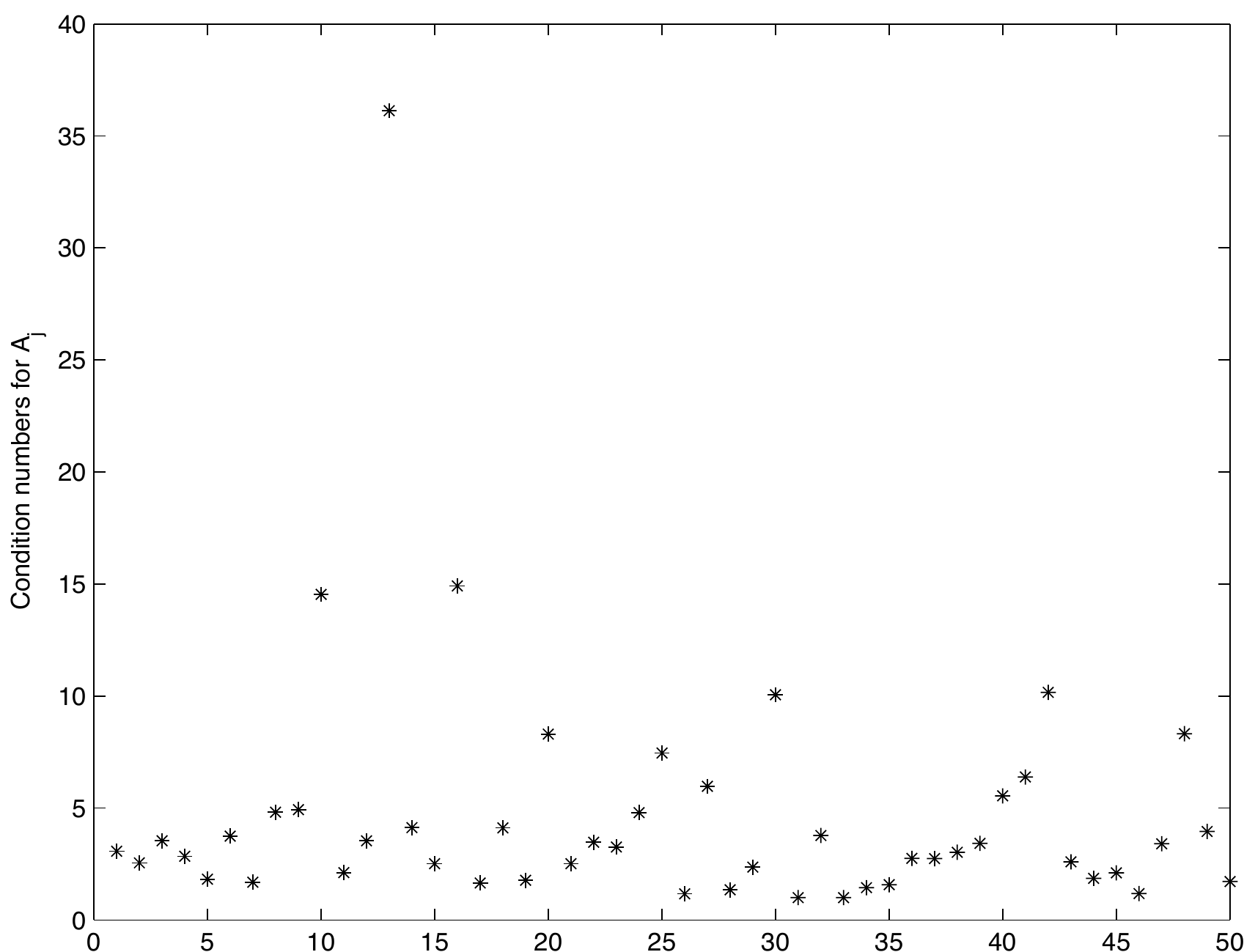}  & \includegraphics[scale=0.43]{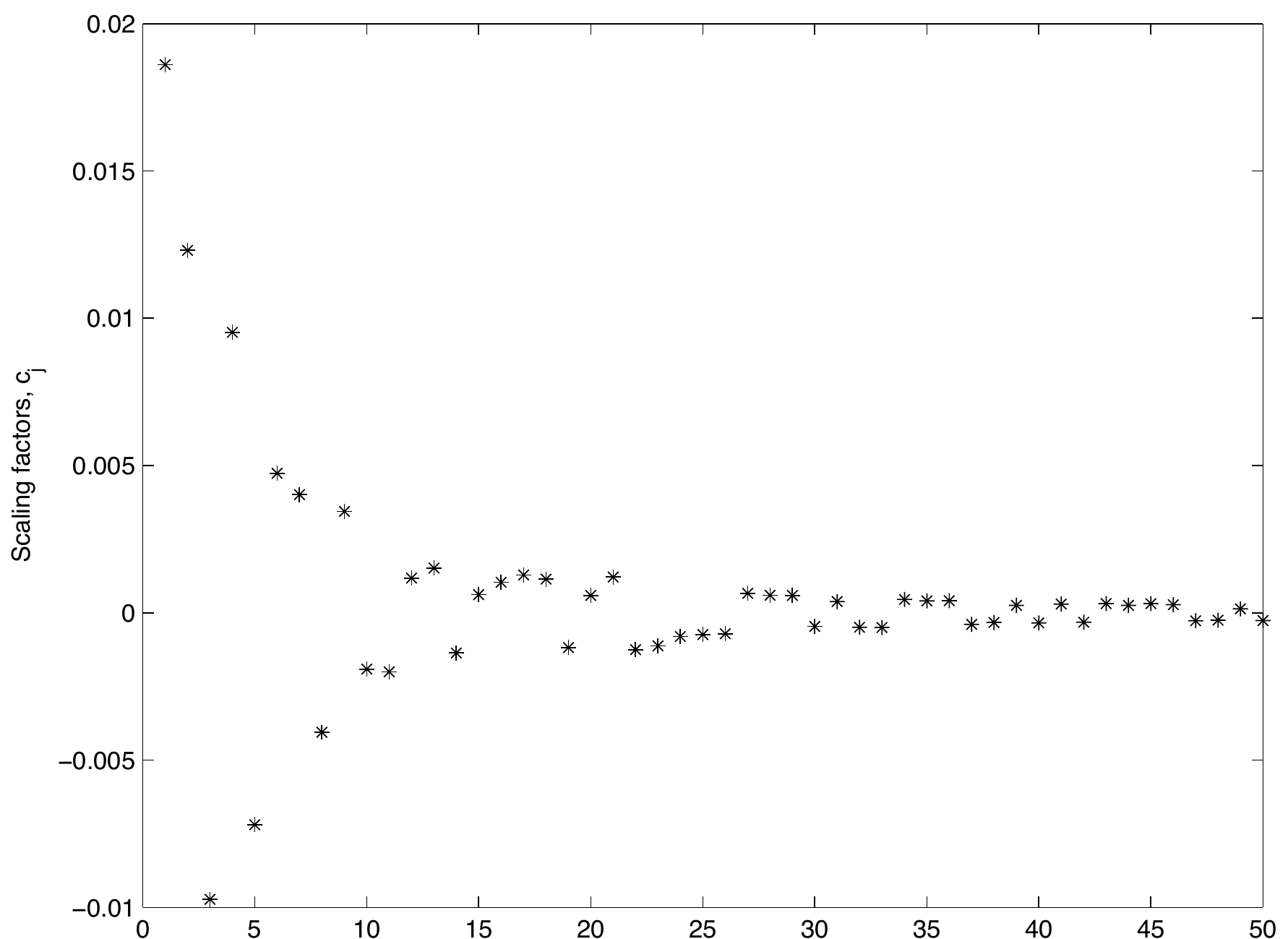}
\end{tabular}
\caption{\label{figs:scaling} A typical evolution of the condition numbers of the shape parameters (left) and the scaling factors $c_j$ (right), plotted agains the number $j$ of the function in \eqref{E}.}
\end{figure}

Another indication of a consistent behavior of the iterative algorithm proposed here is the evolution of the parameters computed dynamically in each iteration. Although the eigenvalues of a shape matrix $A_j$ tend to grow and can become eventually large (when fitting a small and steep area), their ratio (the spectral condition number of $A_j$) remains bounded with some exceptions; recall that the condition number 1 corresponds to an (isotropic) radial basis function. On the other hand, the scaling factors $c_j$ steadily decrease, in concordance with a  gradual reduction of the residual errors, see Figure~\ref{figs:scaling}.

Regarding the stopping time, the experiments performed with real and synthetic corneas show that the reasonable number of iterations lies between 20 and 40; there is no clear correlation between the number of iterations and the condition of the cornea, as Figure~\ref{figs:Zernike} shows. This is why we consider appropriate to perform 50 iterations (taking advantage of the speed of the algorithm) in order to decide the right value for $n$ in \eqref{E}.

\begin{figure}   \centering
\begin{tabular}{cc}  
\includegraphics[scale=0.5]{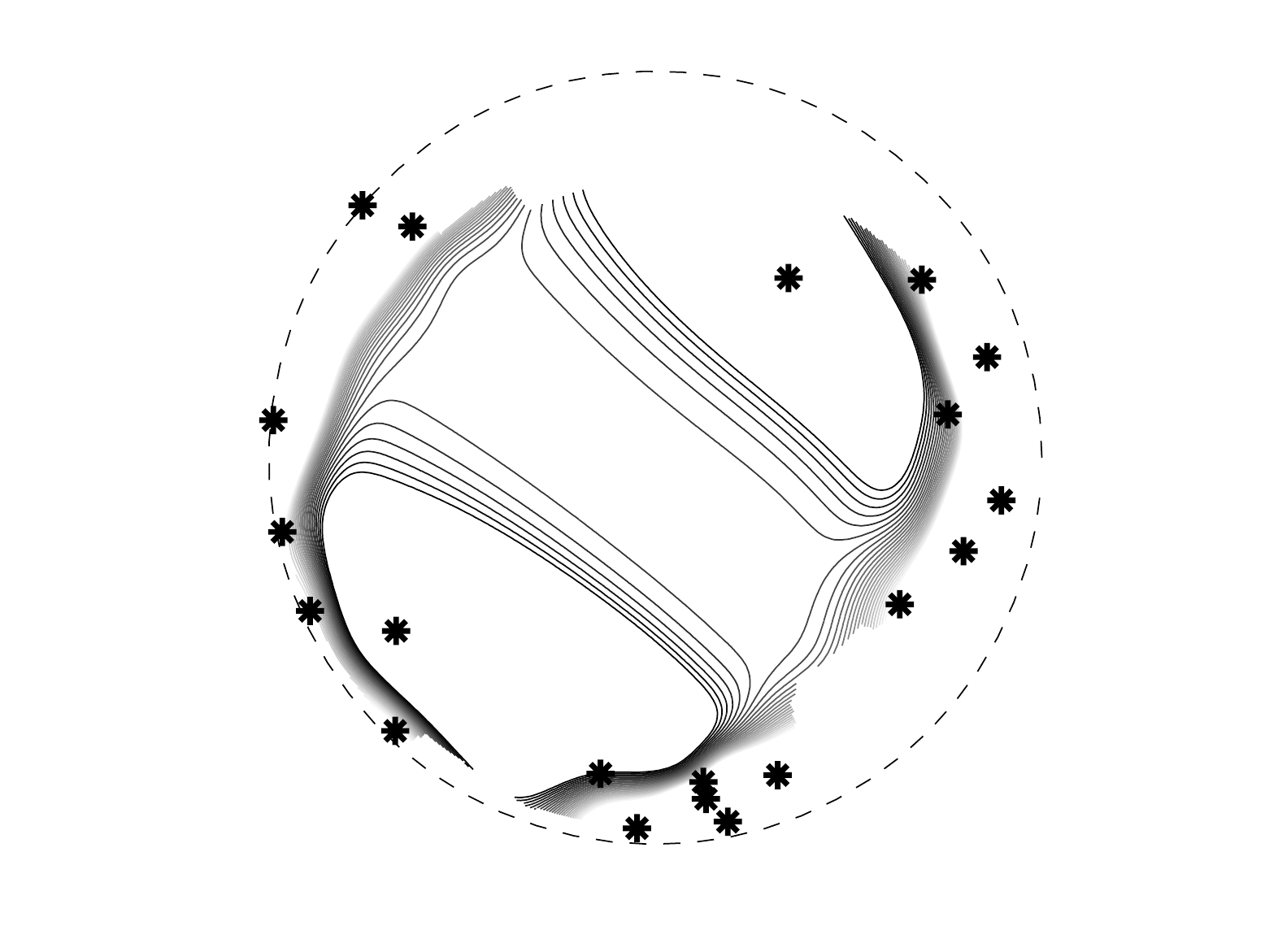}  & \includegraphics[scale=0.5]{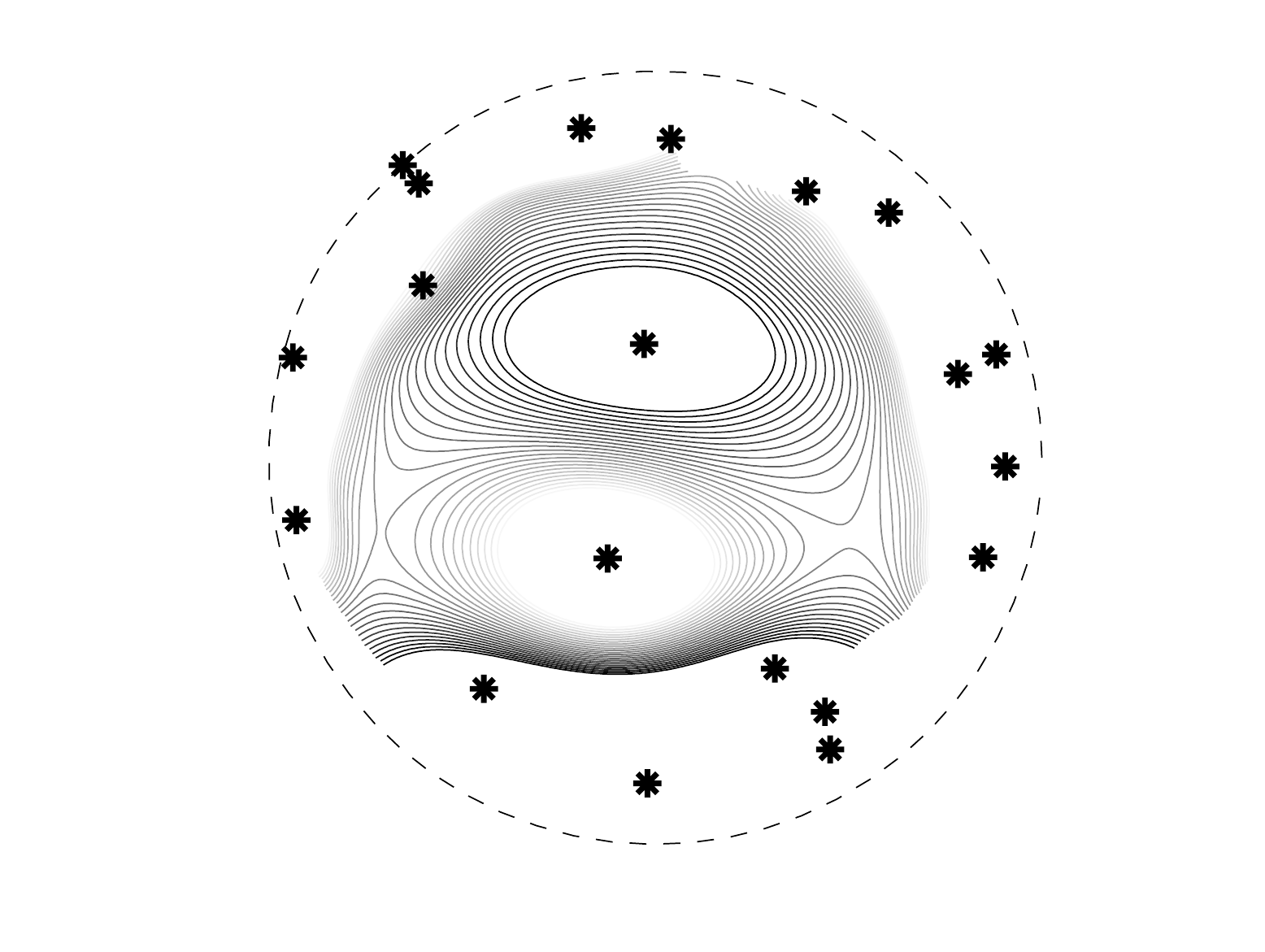}
\end{tabular}
\caption{\label{figs:centers} Center location for the reconstruction with 20 A-RBGF of a normal cornea (left) and a keratoconic one (right). The contours represent the level curves of the residues of the altimetric data with respect to the best fit sphere.}
\end{figure}

However, more correlation exists with the location and grouping of the centers $\QQ^{(j)}$: for the normal corneas the centers typically cluster at the border of the area, where most of the oscillations occur, while for corneas affected by keratoconus we observe how some centers match the deformation already at the first iterations, see Figure~\ref{figs:centers}.

\begin{figure}   \centering
\includegraphics[scale=0.5]{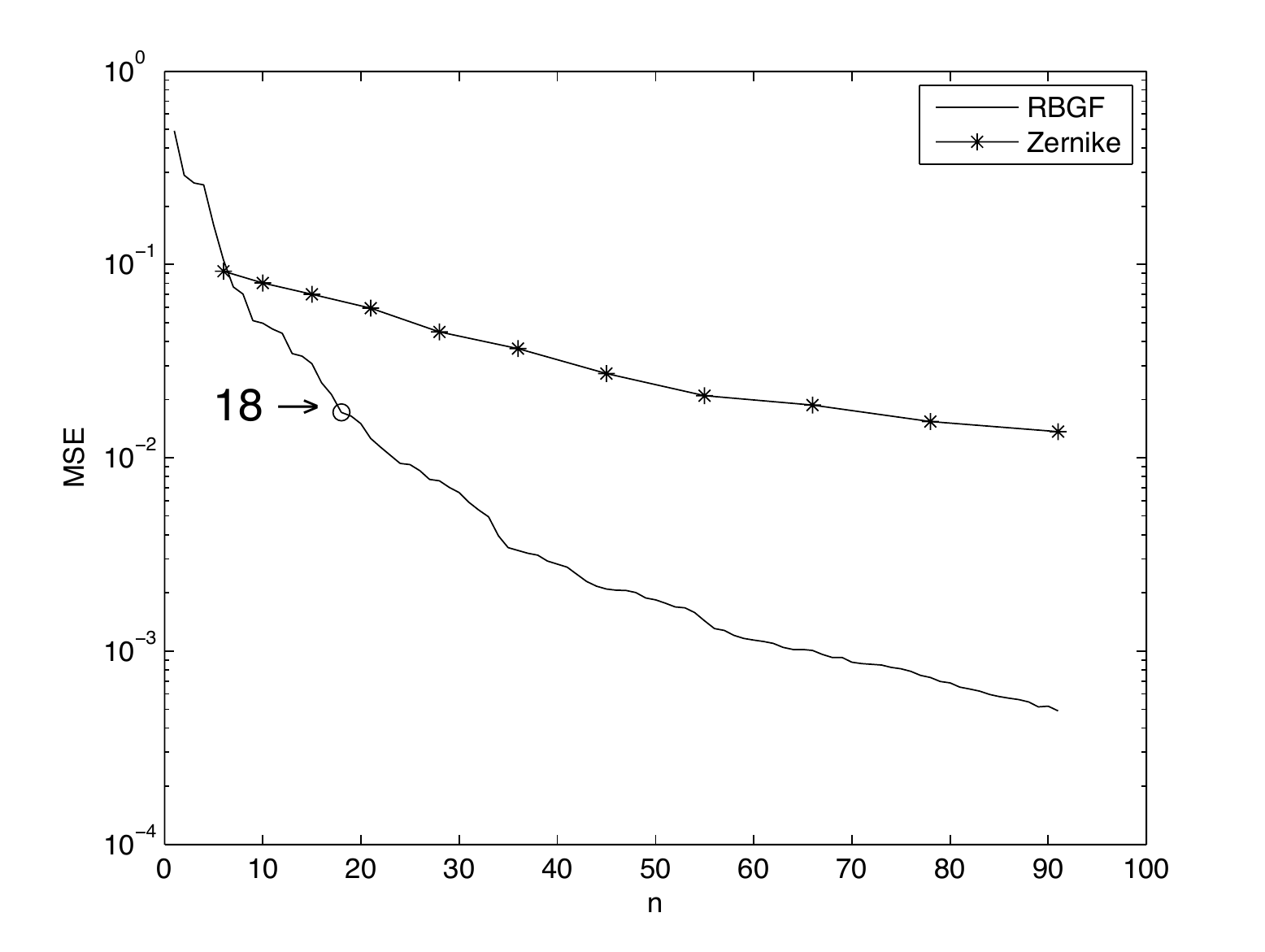}  %
\caption{\label{figs:power} The $MSE_j$ for corneal power data reconstruction by Zernike polynomials and by the A-RBGF algorithm.}
\end{figure}

The adaptive algorithm described here is suitable for a reliable reconstruction of any surface for the discrete set of data. In particular, we can also reconstruct a corneal power map or the wavefront, see an example in Figure~\ref{figs:power}. Taking into account the typical shape of such a surface, it is convenient here to skip the fit by a sphere or Zernike polynomials of a low order, making $S\equiv 0$ in \eqref{analyticExp}.

\begin{figure*}   
\begin{tabular}{cc}  
\includegraphics[scale=0.45]{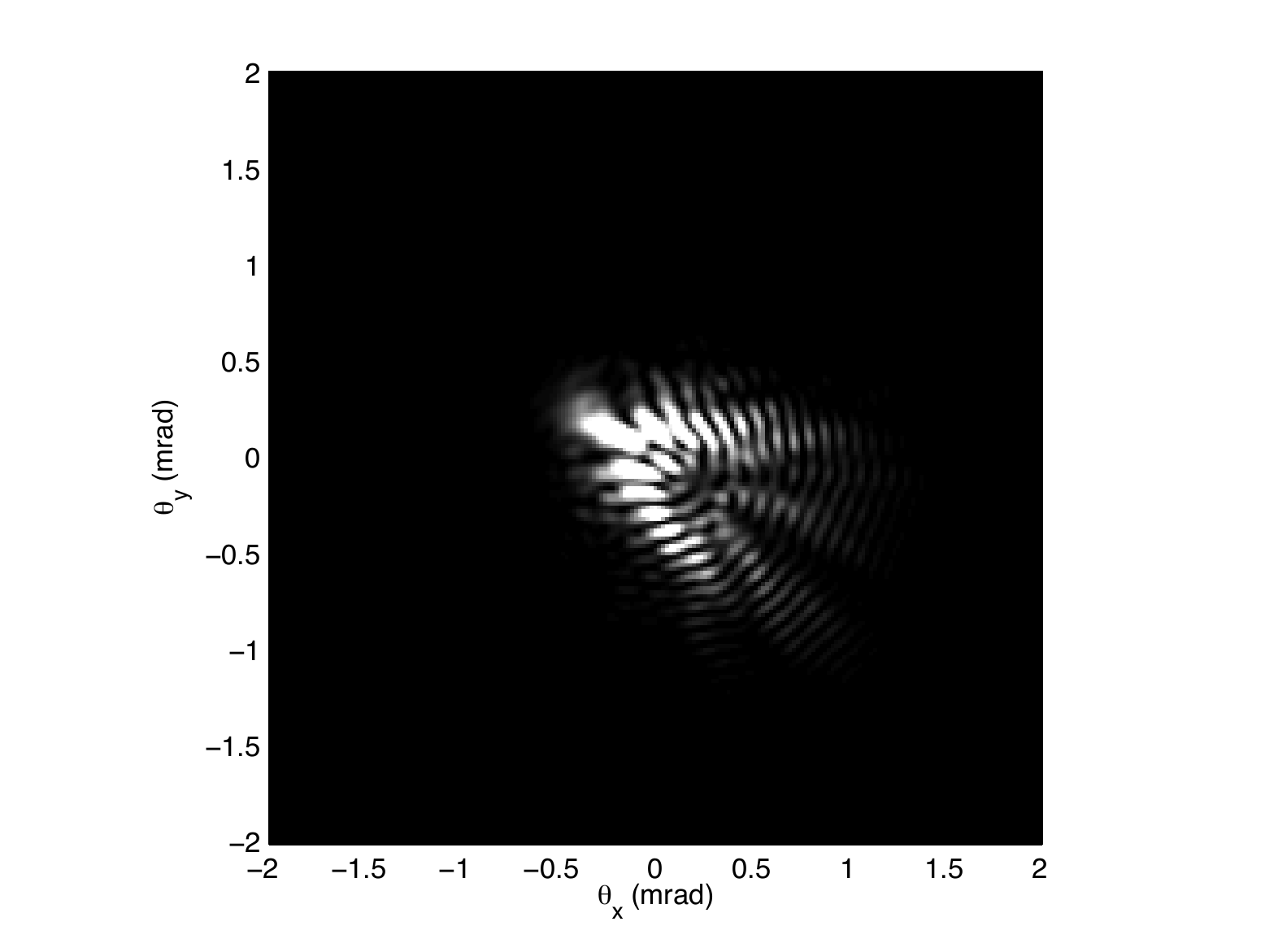} & \includegraphics[scale=0.45]{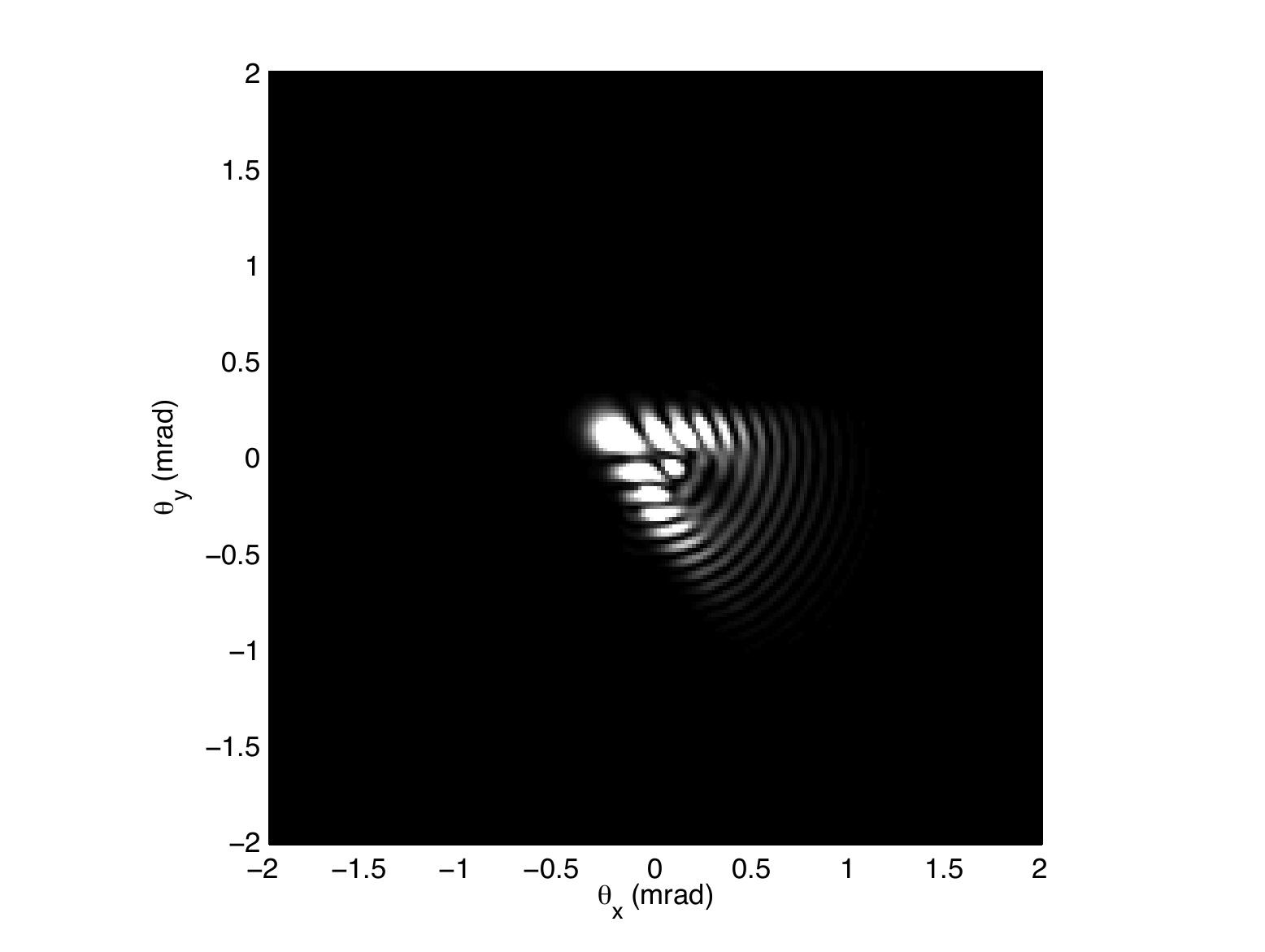} \\
\includegraphics[scale=0.45]{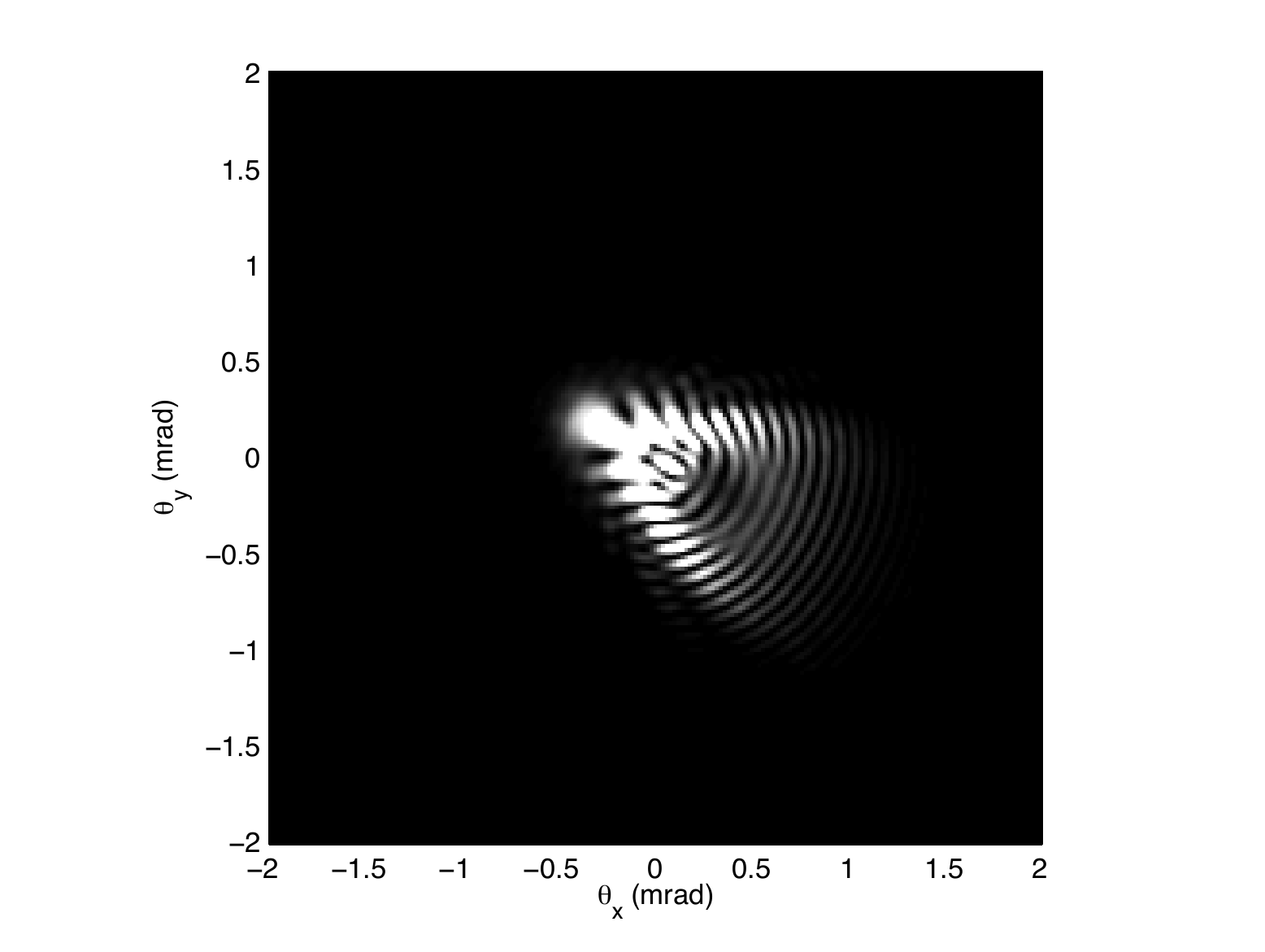} & \includegraphics[scale=0.45]{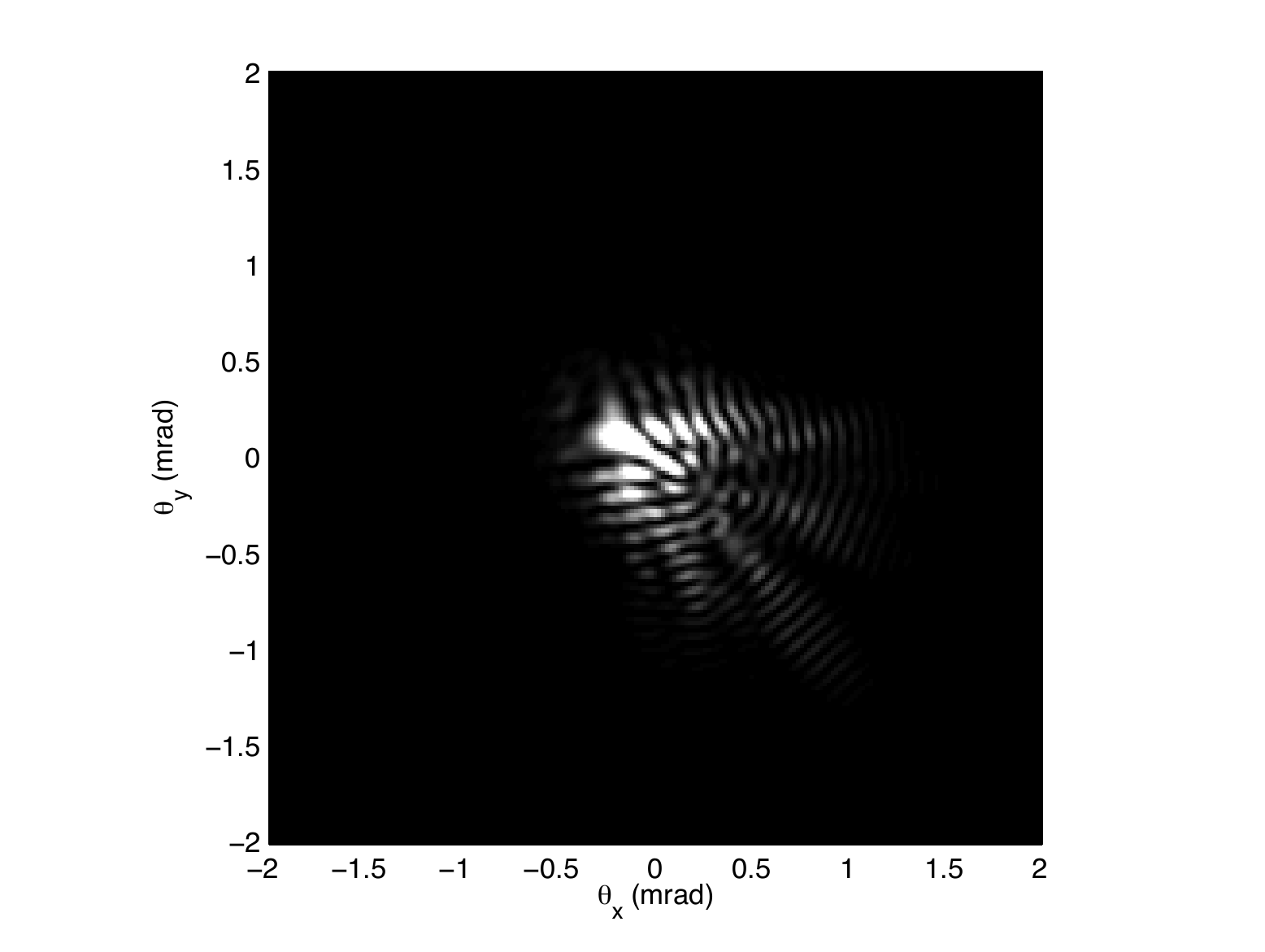}
\end{tabular} 
\caption{\label{figs:PSF} Original PSF for a simulated cornea with a pupil size of 8 mm of diameter and high wavefront error (upper left), and the corresponding PSFs from a reconstruction with first 21 (order $\leq 5$) Zernike polynomials (upper right), first 55 (order $\leq 9$) Zernike polynomials (lower left), and with 21 A-RBGF functions (lower right).}
\end{figure*}

Although the main goal of the algorithm proposed here is the reconstruction of the topography of the anterior surface of the cornea, it is convenient to assess the quality of the approximation by analyzing the resulting point spread function (PSF). Fig.~\ref{figs:PSF} shows the effect of modeling the corneal surface data of a simulated highly irregular eye on the estimated PSF. The original PSF corresponding to a wavefront described by an expansion in 136 Zernike polynomials and a corneal diameter of 8 mm is shown together with a  Zernike polynomial approximation of radial order $\leq  5$ and $\leq  9$ ($21$ and $55$ functions, respectively), and the A-RBGF approximation with 21 functions. Clearly, the latter leads to a PSF that more closely resembles the original PSF (capturing some finer features) than that derived from the Zernike polynomials.

\section{Discussion and summary}

In this work, we develop an adaptive fitting method for corneal data,  combining modal simplicity with advantages of zonal reconstruction. It consists of a preliminary fit of the data with some global function (sphere or a combination of Zernike polynomials of a low order) and an iterative procedure that adds terms to the analytic representation of the corneal data. Each term consists of a scaled anisotropic radial basis gaussian function. The coefficients are computed dynamically and allow to fit the data in each iteration independently of the scale. The method comprises also a filtering procedure that discards the outliers (data clearly corresponding to measurement noise) and a stopping criterium that chooses the final number of functions in the analytic representation in accordance with the evolution of the residual error.

The numerical implementation of this algorithm in a standard personal computer is very fast (execution time below 2 seconds). Experimental results allow us to draw the following conclusions: 
\begin{itemize}
\item the least square approximation by a linear combination of Zernike polynomials of a radial order up to 6 (which is the standard in modern aberrometers \cite{Rozema2005}) fits adequately the altimetric data in the case of a normal cornea. It can be used also to capture the major features of the shape of the surface. However,  for strongly aberrated corneas the Zernike-based procedure saturates relatively early, and we need a high number of terms to achieve the desired accuracy at regions of localized steepening, at the price of overparametrizing the model and fitting the measurement noise. Last but not least, the complexity of each individual term in the functional representation increases with its index.
\item in contrast, the iterative method presented here exhibits a steady exponential error decay, independently of complexity of the cornea. Its actual rate is basically influenced by the residue distribution: the fast decrease in the first iterations, when all salient features are reconstructed, is followed by a stable linear decay, when essentially errors are being fit. This can be used as a stopping criterium for the iterations. In this way, the minimal amount of functions in the analytic representation of the cornea is used.  
\item unlike in the case of Zernike polynomials, the complexity of each term in the functional representation with A-RBGF is the same, but their parameters vary to fit the current scale. This scale is determined only by the residual errors and not by the number of the iteration. 
\item due to the localized character of A-RBGF, the position and clustering of their centers, as well as the size of the shape parameters, provides an additional spatial information about the regions of higher irregularity. These ideas were actually used in elaboration of new cornea irregularity indices that are currently under study.
\item Zernike-based reconstruction, being center-oriented, is also very sensitive to rings with incomplete data. In clinical practice this is usually motivated by eye lashes obstruction or a tear film disruption. Still, a large portion of data of each incomplete mire is available and used by the iterative reconstruction with the A-RBGF.
\end{itemize}

The iterative adaptive algorithm for the cornea modeling proposed here provides a method of obtaining a compact mathematical description of the shape or power map of the cornea. All information is ultimately encoded in the following set of values: center and radius of the best-fit sphere, plus the center locations, shape parameters and scaling factors. This description can be used for global visualization of the cornea or of its portions, capturing smaller details than with standard procedures. It can serve also as the input data for resampling and computation of some other relevant values via ray tracing, numerical integration and others. 

\section*{Acknowledgments} 

A.M.-F. is supported in part by Junta de Andaluc\'{\i}a grants FQM-229 and P09-FQM-4643, and by the Ministry of Science and Innovation
of Spain (project code MTM2008-06689-C02-01). Both A.M.-F. and D.R.-L. are also supported in part by Junta de Andaluc\'{\i}a grant 
P06-FQM-01735.

\section*{Bibliography}


\def\cprime{$'$} \def\cprime{$'$}

\end{document}